\documentclass[11pt,preprint]{aastex}
\usepackage{graphicx}

\shorttitle{{\it XMM-Newton} ULX Study}
\shortauthors{Winter et al.}

\begin{document}
\title{{\it XMM-Newton} Archival Study of the ULX Population in Nearby Galaxies}

\author{Lisa M. Winter}
\affil{ Astronomy Department, University of Maryland, College Park, MD 20742}
\email{lwinter@astro.umd.edu}

\author{Richard F. Mushotzky}
\affil{Goddard Space Flight Center, Greenbelt, MD 20771}
\email{richard@milkyway.gsfc.nasa.gov}

\author{Christopher S. Reynolds}
\affil{ Astronomy Department, University of Maryland, College Park, MD 20742}
\email{chris@astro.umd.edu}

\begin{abstract}
We present the results of an archival {\it XMM-Newton} study of the bright
X-ray point sources (L$_X > 10^{38}$\,erg\,s$^{-1}$)
 in 32 nearby galaxies. From our list of
approximately 100 point sources, we attempt to determine if there is
a low-state counterpart to the Ultraluminous X-ray
(ULX) population, searching for a soft-hard state dichotomy similar to that known
for Galactic X-ray binaries
and testing the specific predictions of the IMBH
hypothesis.  To this end, we searched for ``low-state" objects, which we defined
as objects within our sample which had a  spectrum well fit by a simple 
absorbed power law, and ``high-state" objects, which we defined as objects
better fit by a combined blackbody and a power law.  Assuming that ``low-state''
objects accrete at approximately 10\% of the Eddington luminosity \citep{don03}
and that ``high-state" objects accrete near the Eddington luminosity we further
divided our sample of sources into low and high state ULX sources.  We classify 16 sources as low-state ULXs and 26 objects as high-state ULXs.   As in Galactic 
black hole systems, the spectral
indices, $\Gamma$, of the low-state objects, as well as the luminosities, tend to
be lower than those of the high-state objects.  The observed range of
blackbody temperatures for the high state is 0.1-1\,keV, with the most luminous systems
tending toward the lowest temperatures.  We therefore divide our high-state ULXs
into candidate IMBHs (with blackbody temperatures of approximately 0.1\,keV)
and candidate stellar mass BHs (with blackbody temperatures of
approximately 1.0\,keV).   A subset of the candidate stellar mass BHs have spectra
that are well-fit by a Comptonization model, a property similar of Galactic BHs
radiating in the ``very-high" state near the Eddington limit.
\end{abstract}

\keywords{galaxies: general --- surveys --- X-rays:binaries --- accretion, accretion discs}

\section{Introduction}
Through X-ray observations of nearby galaxies, a class of
Ultraluminous X-ray (ULX) sources has emerged.  These are pointlike,
non-nuclear sources with observed X-ray luminosities greater than 
$10^{39}$\,erg\,s$^{-1}$ \citep{mil04}.
Of most interest are those sources with bolometric luminosities in excess of the
Eddington limit for a 20 M$_{\sun}$ black hole, or L$_{bol} > 2.8
\times 10^{39}$\,erg\,s$^{-1}$.  The true nature of these sources is
unclear, and this class most likely includes several different types
of objects.  Though some of these sources are located within a few
parsecs of their host galaxy's dynamical center, they do not exhibit
many of the characteristics of active galactic nuclei (AGN).  Because
the ratio of X-ray to optical flux is a factor of 10 greater
than that of AGN \citep{and03,sto83}, these objects are fairly easy to
recognize in X-ray imaging data.

Assuming that the Eddington limit is obeyed by black hole accretion, the existence of such 
luminous non-AGN sources presents a puzzle. 
Several models have been proposed to account for the high luminosities of the ULXs.  
Among these are relativistic and non-relativistic beaming 
from stellar-mass black hole systems \citep{rey97, kin01, kord02} and accretion of matter into intermediate mass black holes (IMBHs).  
In several ULX systems 
(NGC 1313 X-2, M81 X-9, etc.), detection of emission nebulae surrounding the ULX supports isotropic emission from the central source \citep{pak03}, 
which cannot be described through beaming.  Further, a number of ULX (NGC1313 X-1, etc.) X-ray spectra are best fit with combined 
multi-component blackbody (MCD) and power law fits, similar to Galactic black holes in their high-state.  Recently, \citet{mill04} find that 
many spectral fits of ULXs require cool accretion disk temperatures of approximately 100\,eV. The theoretical relationship between black hole 
mass and disk temperature (T $\propto M^{-1/4}$) has been observed to hold true for stellar mass (typically around 1\,keV) and supermassive 
(around 10-100\,eV) black holes \citep{mak00, por04, gie04}. 
Using these scaling relations, the cool accretion disk 
ULXs would correspond to a population of high-state IMBHs with masses of $\approx 16 - 10^4$\,M$_{\sun}$.

If ULXs do not obey the Eddington limit, they could be the result of an outburst 
(such as can occur in low mass X-ray binaries within our own Galaxy). 
\citet{jon04} find evidence for approximately 5 Galactic black hole X-ray binaries
which exhibit luminosities in the ULX range during outbursts.  These sources would appear as
transient ULXs.  The typical time scale for outburst of Galactic X-ray transients is
a few days to rise from quiescent level with a decline from peak brightness to quiescent value
of 30 - 40 days \citep{che97}.
Another possible explanation is super-Eddington emission from accretion disks 
surrounding stellar mass black holes \citep{beg02, ebi03}.  Sources of this type would be
expected to have soft X-ray components well modeled by hot accretion disks ($\approx 1.3-2.0$\,keV)
similar to superluminal X-ray sources in the Galaxy (e.g. Belloni et al. 1997).

Likely, ULXs include a variety of different objects with both isotropic and non-isotropic emitters.
However, if some ULXs do indeed represent a class of high-state IMBHs, similar
to the high/soft (thermal dominated) state stellar mass black holes in our galaxy, we might
also expect to see the low-state objects from this same population.
In Galactic black hole systems, the low-state is generally
characterized by lower luminosity, with L$< 0.1$\,L$_{Edd}$
\citep{don03}, and a power law photon spectrum, typically with index $\Gamma
\approx 1.7$ \citep{mcc04}.  Indeed, the existence of some ULX sources 
(IC 342 X-1, NGC 5204 X-1)
as possible low-hard (pure power law) state IMBHs, well-fit
by simple absorbed power laws, have been noted from Chandra observations by
\citet{rob04}.
In this study we seek to test a direct prediction of the
IMBH hypothesis; namely, whether there is a class of sources with properties consistent
with what we expect of low-state IMBHs.  This requires two major assumptions:
(1) that the emission from ULXs is isotropic and (2) that IMBHs exhibit the same states
(whose classification was based on luminosity and spectral form) as stellar mass black
holes.    Our goal is to 
find these ``low-state'' sources, if they exist, classify the properties of both 
high-state and low-state ULXs, and test whether these data are consistent or
inconsistent with the predictions of the IMBH hypothesis. 

We present the results of a detailed analysis of ULXs in nearby
galaxies observed with the European Space Agency's {\it XMM-Newton}
observatory.  Only {\it XMM-Newton} provides the count rates and
bandpass necessary to distinguish different spectral models for most
ULXs, accurately determine both the temperature of the thermal
component expected for high-state objects, and determine whether this component
is required in the spectral modeling of these objects.  Since the XMM-Newton
X-ray spectra of ULXs are similar in quality to spectra for Galactic X-ray binaries obtained
in the 1980s, our spectral classification in this paper will remain purely schematic.  Thus,
our classifications as low and high state objects are a first approximation, based on the
quality of the spectra available.
  
In Section 2, we detail the observations examined from the {\it
XMM-Newton} archives and explain the data analysis for the individual
point sources. In Section 3, we discuss the spectral fitting
technique as well as simulations we conducted to determine their
validity.  We discuss the implications of our results in Section 4.

\section{Observations and Data Reduction}

The data used in this investigation were drawn from the {\it
XMM-Newton} public data archive.  Assuming that
low-state ULXs exist in the luminosity range of
{$10^{38-39}$\,erg\,s$^{-1}$}, we conducted simulations to determine
the optimum criteria for observations capable of resolving point
sources of this luminosity. Our simulations provided a guide
for choosing which of the vast number of archival XMM datasets
we should examine.  This luminosity range was chosen on the
assumption that an approximately 100\,M$_{\sun}$ black hole would
radiate at $\approx 10\%$ of the  L$_{Edd}$ in the low-state
\citep{don03}. 

Within the luminosity range of interest (L$_X > 10^{38}$\,erg\,s$^{-1}$),
there are a number of known objects that could be confused with ULX sources.
One type of source is supernova remnants (SNRs).  These sources are often easy to distinguish
based on their characteristic spectrum: with poor signal to noise we
expect a steep power law and as the signal to noise increases, emission
lines become clearly visible.  Super-Eddington accreting neutron stars have been
observed to have luminosities within this range for a short period of time.  
Neutron star X-ray binaries often have spectra well fit by a
hot multi-color disk blackbody model, or with low signal to noise, by a bremsstrahlung model.
Both models have similar curvature and a 0.7-2.0\,keV blackbody model
is indistinguishable from the bremsstrahlung model.
  We chose to use the bremsstrahlung model 
because it is the simpler model and gives an adequate qualitative description to the data.
We expect that for low temperature bremsstrahlung sources, the spectrum should be
easily distinguishable from a power law with $\Gamma \approx 1.7$ (as is
expected for a low-state object).  If, however, the neutron star spectrum
has kT $> 5$\,keV, as observed for some NS X-ray binaries, our simulations show 
that we can not
distinguish between the power law and bremsstrahlung models.

The most common sources we expected to find in this luminosity range were
the analogs to Galactic black hole X-ray binaries in their high-soft (thermal
dominated) state.
These sources typically have spectra well fit by a blackbody with temperature
of $\approx 1.0$\,keV combined with a power law with index $\Gamma \approx 2.5$.
Our simulations sought to determine the number of photons
required to distinguish between spectral fits corresponding to a power law
model with $\Gamma \approx 1.7$ and a combined blackbody and power law model.
These models qualitatively correspond to those of a low-state (pure power law spectrum)
and high-state (thermal dominated spectrum)
X-ray binary. Since we do not know the proper normalization between the blackbody
and power law components for high-state objects (it varies from source to source),
we tested whether each of the components separately, e.g. blackbody or a steep power law,
 could be distinguished from
the simulated ``low-state'' spectrum. We chose to simulate spectra in {\it XSPEC} using the
command {\it fakeit none}.  We used generic response and ancillary response
matrices.  Simulating a power law model with a $\Gamma = 1.7$, we found that
for 200, 400, and 1000 counts, these models were distinguishable at $> 99$\% confidence from a
blackbody source (with kT constrained to the range of 0.6 to 1.3\,keV, similar to that
of Galactic black holes).  We
found that for a lower number of counts the distribution in $\Gamma$ values increases to 
include a larger range of $\Gamma$ values (i.e. $\Gamma$ = 1.3 - 2.0 compared to
$\Gamma$ = 1.5 - 1.7).  
Simulating a power law with $\Gamma = 2.5$, we find the
same trend.  We determined that at roughly 400 counts the distributions of $\Gamma$
from a $\Gamma = 1.7$ and $\Gamma = 2.5$ power law 
become entirely separable at $> 99$\,\% confidence. 

In order to distinguish between the different spectral fits
for objects with L$_X \sim 2 \times 10^{38}$\,erg\,s$^{-1}$, we select all
 galaxies that were observed for at least 10 ks (with the
exception of the bright ULX in NGC 5408, which had enough photons for
analysis despite the low exposure time) with {\it XMM-Newton} and that
are closer than 8 Mpc.  We estimated that these criteria would
give us a minimum of 400 counts for objects with L$_{X} > 2 \times 10^{38}$\,erg\,s$^{-1}$.
We emphasize that the criteria quoted, based on the simulations, were used as a 
guide in choosing the sample of galaxies examined in this study.  These simulations 
are not used as the statistical basis for our object-by-object analysis 
(discussed in Section 3).

Our sample of galaxies is selective in that it represents objects of
interest in the X-ray band.  We include details on these host galaxies
in Table~\ref{tbl-1}.  {\it XMM-Newton} spectral information of individual
X-ray sources had previously been published for approximately 60\%
of the host galaxies.  We include references in the alternate ID column
and footnotes of Table~\ref{tbl-2}.  We do not compare our results with
these previous studies on a source by source basis.   

\begin{figure*}
\plotone{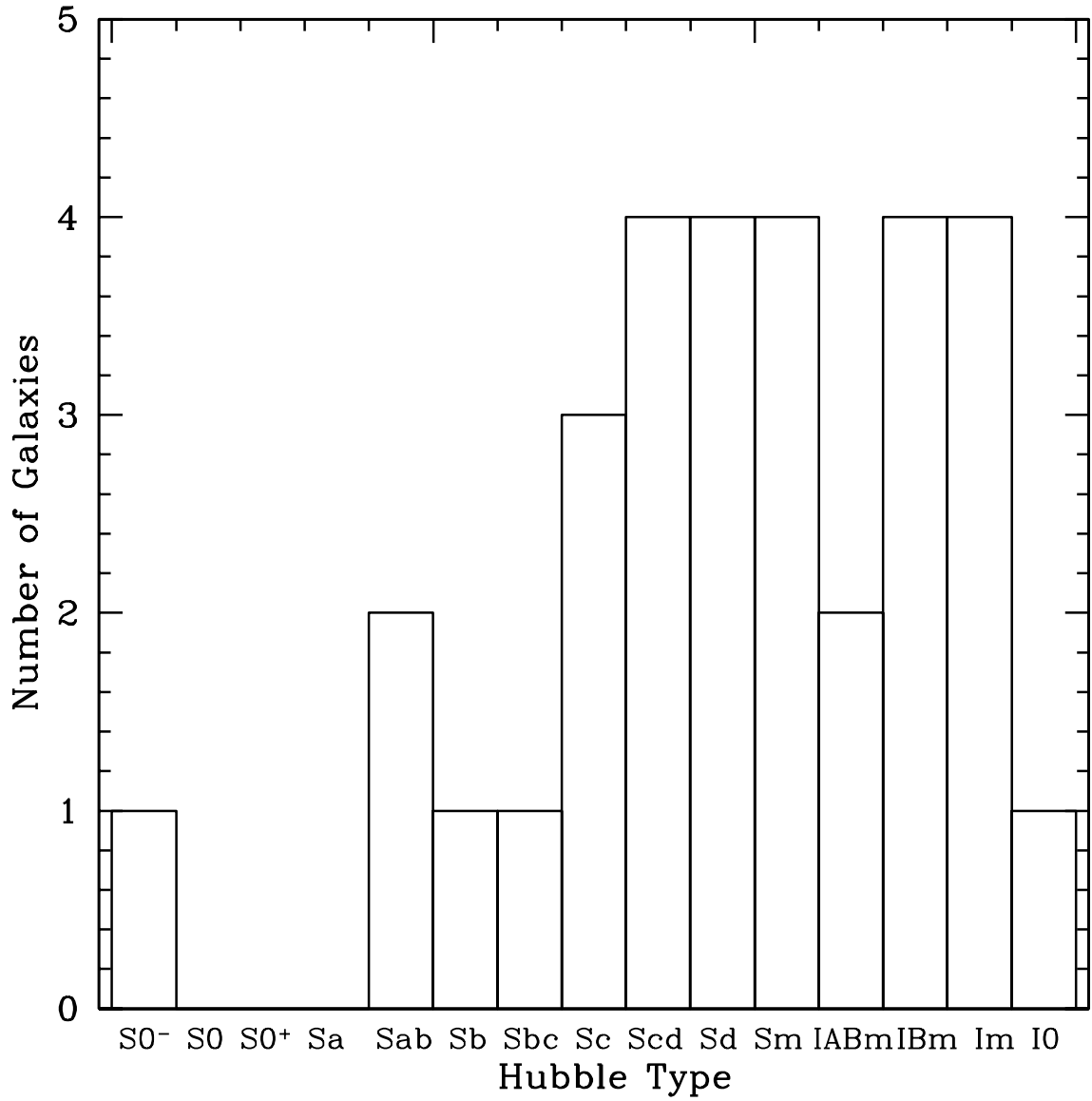}
\caption{Distribution of galaxies by Hubble type among our archival
{\it XMM-Newton} sample of nearby ($< 8$\,Mpc) galaxies.  Our sample
consists solely of spirals and irregulars.\label{fig1}}
\end{figure*}

We found that abstracts describing the proposals for {\it XMM-Newton} observations
were available for only 13 of the 32 galaxies examined.  Of these 13, only one
observation cited the motive as a study of ULXs (NGC 1313).  However, 7 of the
remaining 19 galaxies contained sources classified as IXOs, intermediate X-ray
objects, by \citet{col02}.  If the remaining galaxies were not studied due to
their ULX population, the effects of bias are small with roughly 25\% of the sources
studied explicitly due to their connection with ULX sources. Our host galaxies include 
only spirals and irregulars.  Figure~\ref{fig1} displays 
the distribution of galaxy type.

We reduced the data using the {\it XMM-Newton} Science
Analysis System (SAS) version 6.0.0.  Since the processed pipeline
products (PPS) were created with earlier versions of SAS, the
observation data files (ODF) were used to produce calibrated photon event
files for the EPIC-MOS and PN cameras using the commands {\it emchain}
and {\it epchain}.  Following this, the events tables were filtered
using the standard criteria outlined in the {\it XMM ABC Guide}.  For
the MOS data (both MOS1 and MOS2 cameras), good events constitute
those with a pulse height in the range of 0.2 to 12 keV and event
patterns that are characterized as 0-12 (single, double, triple, and
quadruple pixel events).  For the PN camera, only patterns of 0-4
(single and double pixel events) are kept, with the energy range for
the pulse height set between 0.2 and 15 keV.  Bad pixels and events
too close to the edges of the CCD chips were rejected using the
stringent selection expression ``FLAG == 0''.  

Time filtering was
applied as needed by editing the light curve produced in {\it
xmmselect} for the entire observation.  Flare events (distinguished by
their high count rate) detected in all three cameras, were cut using
the {\it tabgtigen} task as outlined in the {\it ABC Guide}.  Typical
count rate parameters for filtering were 'RATE $< 5$' for MOS detectors
and 'RATE $< 20$' for the PN detector.  Such
filtering was only done as needed.  Pre-filtered exposure times are 
listed in Table~\ref{tbl-1}.  The number of counts from the filtered net exposure times for the
individual sources are listed in Table~\ref{tbl-2}.  We note that 
the filtered data are not always sufficiently clean that a high signal-to-noise
is maintained up to 10\,keV.  Sources with a high background flux level,
relative to the source spectrum, show poorer signal-to-noise in the spectrum above 1\,keV.

Before extracting spectra of the brightest sources, contour maps of
the X-ray observation were overlaid on Digital Sky Survey (DSS)
images.  This ensured that bright foreground stars and background AGN
were easily distinguished, and thereby not included in our sample.  
Also, we checked the {\it XMM-Newton} source positions with NED and
SIMBAD to determine if they coincide with any known background
galaxies or QSOs.  A list of these bright fore-ground or background
sources is included in Table~\ref{tbl-7}.

\section{Spectral Fitting}
Spectra for the bright point sources were extracted based on their
apparent brightness in the CCD images.  No explicit source detection algorithm
was necessary.  We used the SAS task
{\it especget}.  With this task we created spectra (for both the
source and background), response matrices, and ancillary response
files for all three EPIC cameras, when possible.  The typical
extraction radius was 20 arcseconds, but depending on both the size
and proximity of a source to another source, the extraction radius ranged from
9 - 87 arcseconds.  Background spectra were extracted
either in an annulus centered on the source, or in a circle of
appropriate size away from the source, depending on the
proximity of the candidate source to other X-ray sources.  Annuli were used
for sources that were not located within a few arcseconds of another source,
thus annular background extraction radii were not used for sources with small
extraction radii.  For sources in crowded regions, we used circular extraction radii
close to the source. We extracted backgrounds close to the source in
order to correct for emission local to the ULX.  Once the
spectra were obtained, they were rebinned to require at least 20
counts per bin, using the command {\it grppha} in LHEASOFT.  The list
of sources, with position and count information, is included in
Table~\ref{tbl-2}.  We only included sources for our spectral studies
that had 400 or greater PN counts (or MOS for sources in NGC 253 and M81
and NGC 4945 XMM3 and NGC 2403 XMM1, for which PN spectra were not available)..

The extracted spectra were fit with standard models in XSPEC v11.3.1.
For each source, we fit the PN and MOS spectra simultaneously in the
0.3-10 keV range.  We allowed a free normalization constant to account
for the differences in flux calibration between the three cameras
(similar to \citet{jen04}). Each source was first fit with an absorbed
single component model. In all cases we used the standard absorption model {\it
wabs}, leaving the column density as a free parameter.  For those sources
where the hydrogen column density was unconstrained, we fixed the value
to the Galactic foreground value listed in Table~\ref{tbl-1}.  

\subsection{Single-Component Sources}

Results of
the single-component fits are seen in Table~\ref{tbl-3}.  We include
in this table only the best-fit parameters for those sources
best described by a single-component model.    The addition of a blackbody
component to these single-component fits changes the $\chi^2$ value by a negligible
amount and therefore is not statistically significant.  More specifically, 
the addition of a blackbody
model to the power law fit corresponds to a $\Delta\chi^2 < 2.3 $, 
which is the $\approx 68$\% confidence 
level using the F-test for two degrees of freedom.

The flux values quoted
represent the unabsorbed flux in the PN spectra, in the 0.3-10 keV
band.  All errors quoted, here and subsequently, correspond to the
90\% confidence level for one degree of freedom ($\Delta\chi^2 =
2.71$).  The luminosities were calculated from the unabsorbed flux
using the distances quoted in Table~\ref{tbl-1}.  Both flux and
luminosity correspond to those of the best fit model (power law or
bremsstrahlung). It should be noted that since our selection criteria
was based on a count rate cutoff,
due to the variety of spectral forms, the inferred luminosity cutoff will not be uniform.

In Table~\ref{tbl-3} we denote the single component model we choose as the better fit
in bold.  This notation also indicates the model (power law or bremsstrahlung)
used to compute the quoted flux. For $\approx 46$\% of the power law sources, the
$\chi^2$ difference ($< 2$) between the power law and bremsstrahlung models is only
marginally different.  Of these sources, the average kT value for the bremsstrahlung
fit is 5.54\,keV.  From our simulations we find that at high temperatures 
the bremsstrahlung fit becomes indistinguishable
from a simple power law.  Thus, given the high temperatures of the bremsstrahlung
fits for these sources, they are equally well described by the power law model.  
Typical kT bremsstrahlung values for accreting neutron stars are from 3.0 to 
7.0\,keV \citep{jon77}.

\subsubsection{``Low State" ULXs}

From these single-component sources, we classify 16 sources as low-state ULX
sources.  This classification is based on (1) the shape of the spectrum, well-fit
by an absorbed power law,  (2) the luminosity of the sources (they
needed to be luminous enough to be included in our sample, L$_X > 10^{38}$\,erg\,s$^{-1}$),
and  (3) the  X-ray location of the object within the optical galaxy (based on Digital Sky
Survey (DSS) images).   The third criterion was important in limiting the effects of contamination from
fore-ground and background sources within our sample of ULX sources.  Thus, we overlaid
X-ray contours from the XMM image on the DSS images, determining the location of the X-ray source
as within or outside of the optical extent of the galaxy.  We note that for two of the 
sources classified as low-state
ULXs (Holmberg I XMM2 and NGC 2403 XMM4), there is uncertainty of whether the X-ray source
is in fact within the optical galaxy due to the quality of the optical DSS images.
The images used to determine the third
criterion are available online (\url{http://www.astro.umd.edu/$\sim$lwinter/second3.html}).  We further
discuss these criteria in Section 4.

 We state that the first criterion for classification as a low/hard state ULX is a spectrum
well-fit by a power law.  Of the 30 sources in Table 2, three sources were clearly not well-fit
by either the power law or bremsstrahlung model.  Seven of the remaining 27 sources were
clearly not within the optical extent of the host galaxy.  Of the remaining sources excluded
from classification as a low-hard state object IC 2574 XMM1 was better fit with a bremsstrahlung
model (with $\Delta \chi^2 = 13.4$).  In an additional observation of the source NGC 4258 XMM2,
a simple power law model is not an adequate fit to the data (while the luminosity of the source is not
within the ``ULX" classification range mentioned in the introduction).  For the remaining sources,
NGC 247 XMM2 and NGC 253 XMM2, there was sufficient doubt on the spectral form where
the bremsstrahlung and power law model as well as the addition of a thermal component all
yielded adequate fits to the spectra.  Therefore, we excluded these sources from a classification,
noting the ambiguity of the model fits for these sources.

For those sources we classify as ULXs, we include computed bolometric luminosities
in Table~\ref{tbl-11}.  To compute the bolometric luminosities for these ULX
sources, we used the exponentially 
cutoff power law spectrum of \citet{mag95}, model {\it cutoffpl} in {\it XSPEC},
with a cutoff energy of 10\,keV.  
From observational studies of Galactic X-ray binaries, it has been observed that
low-state objects have spectra that cut-off at high energies ($\ga 10-200$\,keV) \citep{zdz04}.  
Thus we chose the exponential model {\it cutoffpl} over a simple power law.  This also
minimizes the total luminosity for flat power law sources.
We computed an unabsorbed bolometric
flux in the $0.1 - 100$\,keV range through use of the {\it dummyresp}
command (which extends the model beyond the observation's energy range).  The
luminosity was then computed using the distances listed in Table~\ref{tbl-1}.
We quote these values as L$_{cutoffpl}$ (the luminosity obtained from extrapolating the power law portion of
the spectrum
as an exponentially cut-off power law) in Table~\ref{tbl-11}.  We note that
these values represent an upper limit on the bolometric luminosity for steep power law
($\Gamma > 2$) objects, since we
would expect the power law component to cutoff at some low energy.  However, for shallow spectrum
($\Gamma < 2$) sources L$_{cutoffpl}$ is a lower limit.  This is because, schematically, a steep
power law diverges at low energies while a shallow power law diverges at high energies.

\subsection{Two-Component Sources}

For a number of sources, we found that an improvement in reduced $\chi^2$ 
 was achieved through fitting their spectra with an
absorbed two-component blackbody and power law model.  We chose a
simple blackbody model over the multi-component disk model, {\it diskbb},
for purely schematic reasons.  Namely, observations of galactic
X-ray binary systems were fit with blackbody models in the 1980s,
when the signal-to-noise of these objects was comparable to that
for our {\it XMM} data for ULX sources.  We also note that the {\it diskbb} model does not give an
entirely accurate physical description of the data as it neglects the
effects of general relativity.  As a schematic model, the blackbody model
is simpler than {\it diskbb}, with the same number of degrees of freedom.
In addition, for low temperatures both models yield virtually identical
temperatures.  For this study, we chose the simpler model.  We defer
to a further paper a discussion of the different models for the
thermal component.    

In Table~\ref{tbl-4} we present the results for the sources which are
fit significantly better by the two-component model, these are sources
where the improvement in $\chi^2$ is greater than 8 (determined
from our simulations in Appendix A).  
We include in Table~\ref{tbl-4} the
improvement in $\chi^2$ of the two-component fit over the simple power
law.  We include the power law best fits to these sources in the
appendix for comparison with other analyses. In order to determine 
whether the blackbody component is
statistically significant for the sources fit with a two-component
model, we simulated spectra based on accurate modeling of some of the
brightest sources: NGC 247 XMM1, NGC 5408 XMM1, and Holmberg II XMM1.
These sources span the observed range of the ratios of the blackbody
to power law component and thus represent those from our sample with a
weak blackbody relative to the power law component, intermediate case,
and a strong blackbody, respectively.  Our simulations are described
in full in Appendix A.  We found that, using a $\Delta\chi^2>8$
criterion, which corresponds to the 99\% significance level as according to the
F-test for the addition of two extra parameters, we can readily detect the strong and intermediate
thermal components in all spectra with more than 400 counts.  The weak
thermal emission cannot be detected in 400 count spectra, but is
readily detected in 2000 count spectra.  This gives us confidence that
our results are statistically meaningful.

\subsubsection{``High-State ULXs"}
Of the sources in Table~\ref{tbl-4}, we classified high-state  (or thermal dominated) ULX
sources  based on three criteria: (1) spectra characterized by an absorbed
power law and blackbody model,  (2) luminosity, and  (3)
X-ray source within the optical extent of the host galaxy.   The luminosity criteria
required that these sources have unabsorbed luminosities L$_X \ga 3\times10^{39}$\,erg\,s$^{-1}$
 (we used L$_X = 2.7\times10^{39}$\,erg\,s$^{-1}$ as our hard cutoff).
If the sources are radiating at the Eddington luminosity, this
cutoff luminosity corresponds to objects with masses greater than 20\,M$_{\sun}$.

 From Table~\ref{tbl-4}, 27 observations are recorded with L$_X > 2.7\times10^{39}$\,erg\,s$^{-1}$.
The addition of a thermal component to these sources is statistically significant over
a pure power law model.  Of the 27 observations, 3 correspond to multiple observations
of a single source.  From an analysis of the DSS images, all 24 of these sources are within
the optical extent of their host galaxies.  However, M51 XMM5 appears to be coincident
with the center of its host, a dwarf companion galaxy to M51.  The location, coupled with
the high luminosity (L$_X = 1.9\times10^{42}$\,erg\,s$^{-1}$) leads us to classify this source
as an AGN.  We also excluded two sources (NGC 1313 XMM2 and M81 XMM2) from our sample 
of high/soft state ULXs due to their previous identification as supernovae.  Of the remaining
21 sources, NGC 253 XMM4 had a luminosity of $2.5\times10^{40}$\,erg\,s$^{-1}$ in one
observation and $2.2\times10^{39}$\,erg\,s$^{-1}$ in a second.  This significant change in luminosity,
with one observation below our luminosity cutoff and another a factor of $\approx 10$ higher
than the other, led us to exclude this source as a high/soft state ULX.  It is likely that this source
is a stellar mass X-ray binary within its host galaxy, where one of the observations captured
the source in an outburst.

In Table~\ref{tbl-13}, we list sources that have $\Delta \chi^2$
values less than 8 for a single observation.  Most of these
sources have weak blackbody normalizations compared to the power law
normalization.  We classify these sources as being well-fit by a
two-component model while acknowledging the uncertainty in the fit as
determined by the simulations.  The addition of the thermal component
is not significant enough for these sources to be classified with
certainty in either Table 2 or 3.  The simple power law fits for these
sources are included with those for sources in Table~\ref{tbl-4} in the
appendix.  We note that due to their high luminosity we included six
of these sources (NGC 4490 XMM2, NGC 4490 XMM3, 
NGC 4736 XMM1, M51 XMM2, M51 XMM6, and M101 XMM3)
with uncertain fit parameters as ULX high-state sources.  Two of these
sources (NGC 4490 XMM3 and M51 XMM6) had unabsorbed luminosities $>
10$ times the $3 \times 10^{39}$\,erg\,s$^{-1}$ cutoff used for
high-state ULX classification.  The other four sources had luminosities
above the threshold, as well as weak blackbody components compared
to the power law (see the appendix for simulations).  We used these
points to justify including these sources with the Table~\ref{tbl-4}
sources in the following discussions with the proviso that their
spectral fits do not indicate absolutely the necessity of the
additional thermal component.  For this reason, we denote these
sources with a special symbol (a circle) in subsequent figures while including
them as ``high-state'' ULX objects.

For our ULX
sources modeled by a combined blackbody and exponentially cutoff power law,
we computed bolometric luminosities using two methods.  The first method
is recorded as L$_{cutoffpl}$ in Table~\ref{tbl-11}.  We computed the 
flux from 0.1 - 100\,keV using an unabsorbed blackbody and exponentially cutoff power law
model using the XSPEC command {\it dummyresp}.  For the second method,
recorded as L$_{bol}$ in Table~\ref{tbl-11}, we estimate a more accurate
bolometric luminosity calculated from the flux in the range
of $2 \times$\,kT\,-\,100\,keV where kT is the blackbody temperature obtained
from the model.  In Galactic X-ray binary systems, the power law component of the
X-ray spectrum is believed to be from Comptonization in a corona.  The photons
supplying this energy originate from the blackbody continuum emanating from the
accretion disk.  Thus, a natural cutoff for this power law component occurs at the
peak emission of the blackbody (which is approximately $3 \times$\,kT).  The estimated
values (obtained from cutting off the combined unabsorbed blackbody and cutoff power law model 
at the value $2 \times$\,kT) differ with regard to the full estimate (flux from the fully integrated blackbody
added to the separate flux from the cutoff power law from
$3 \times$\,kT to 100\,keV) depending on the normalization factors used (for both the
blackbody temperature and the spectral index $\Gamma$).  Choosing three sources displaying
a range of blackbody to power law strength (Holmberg II XMM1, NGC 253 XMM1, and IC 0342 XMM3) we found
that the estimated values were within 88.3, 95.1, and 96.8\% of the more complete estimation.
Given their close proximity (within
approximately 90\%) we quote these estimated values as a good approxmiation to the bolometric
luminosity.   

We note that our bolometric luminosities for all of the classified ULX sources, on average, are a factor of 1.08 
greater than the X-ray luminosities in the $0.3 - 10$\,keV band for the objects best fit
by a combined blackbody and power law.  Thus, to good approximation, the X-ray
luminosity is the bolometric luminosity.  However, for the objects best fit by a
simple power law (low-state ULX sources), the average bolometric luminosity is roughly a factor of 7 greater
than the X-ray luminosity in our band.  This average is dominated by the steep power law
objects, in particular Holmberg II XMM1 ($\Gamma = 3.09$).  Excluding this object, we 
get an average bolometric luminosity that is 2.8 times the X-ray flux and more indicative
of the general properties of these power law-fit objects.

\subsection{Additional Sources}

In addition, in this large sample of point sources, we came across a number of
objects whose spectra were not well fit by the models we
employed.  These sources have luminosities exceeding L$_X \approx 10^{38}$\,erg\,s$^{-1}$,
if they are associated with the host galaxy, and are placed in Tables 2 and 3 as well as Appendix B.  These sources include two
supersoft sources, one possible AGN, and three sources well fit with additional absorption
models (including a partial covering model and a model of hot gas).  
We briefly describe these sources in the appendix.

\section{Discussion}
We have determined best-fit spectral parameters of the bright X-ray
sources in 32 nearby galaxies.  In choosing three ``standard'' models
for our study, we hoped to accurately separate high and low state ULXs
from other types of luminous X-ray sources.  We specifically chose to fit the
data with the bremsstrahlung model in order to identify neutron star
X-ray binaries within our sample.  The models we used are purely schematic,
and they do not physically explain the phenomena occurring, but are standard
and qualitatively simple models often used to fit the spectra of Galactic
X-ray binaries.

We cross-referenced the X-ray
positions of our sources with both NED and SIMBAD in order to identify
known supernovae, galaxies, and stars.  In addition, we examined the
DSS optical images to place the position of our sources within their
respective galaxies.  Such analysis aimed to minimize contamination of
our sample of ULXs with bright background and foreground sources.

\begin{figure*}
\plotone{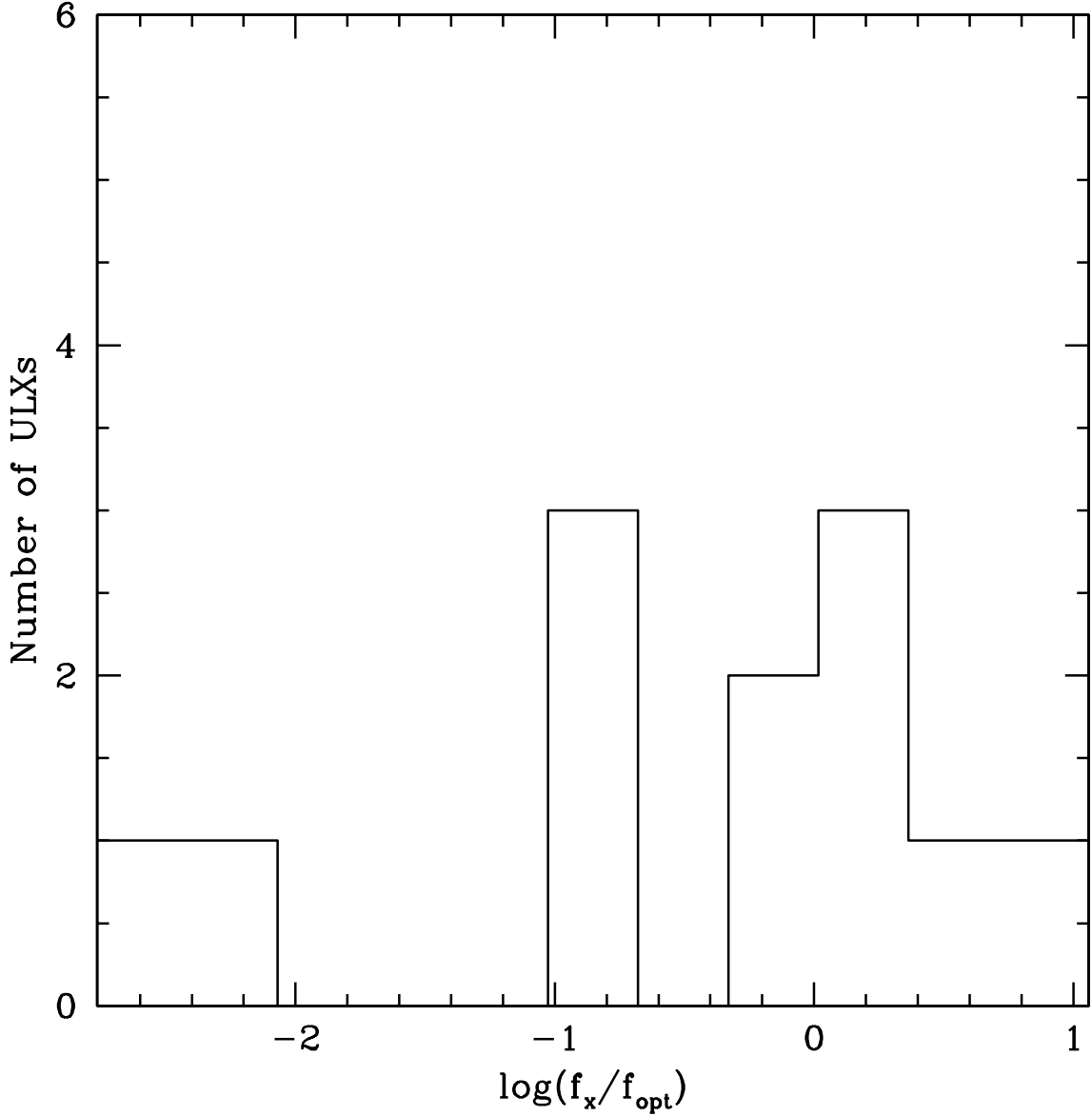}
\caption{The distribution of $f_{x}/f_{opt}$ for the brightest
possible optical point source within the {\it XMM-Newton} error
circle.  We define $f_{x}$ as the unabsorbed X-ray flux in the $0.3 -
10$\,keV range and $f_{opt}$ as the optical flux obtained from the U
filter of {\it XMM}'s OM (as described in text).  These ratios do not
represent the actual $f_{x}/f_{opt}$ of the sources but are an
estimate of the minimum possible value.  A majority of the sources had
no optical point source within the X-ray contour and thus have ratios
of $f_{x}/f_{opt}$ far higher than those indicated in the
plot. \label{fig7}}
\end{figure*}

Further, we examined {\it XMM-Newton's} Optical Monitor data in
the visual bands (U, B, V).  The {\it XMM} PPS contain point source detection
files for the OM data.  We overlaid these point source detections with
X-ray contour maps in order to determine the {\bf brightest possible}
optical count rates for the X-ray sources, which were then converted
into fluxes using the OM calibration documentation.  In
Figure~\ref{fig7}, we plot the distribution of the logarithm of the
X-ray to optical flux for the brightest possible optical counterpart
inside the {\it XMM-Newton} error circle.  Only 13 of the 32 host
galaxies had visible band OM data during the observations.  Of these 13
galaxies, 40 of the X-ray sources were in the range of the OM data and only 14 were
coincident with an optical point source.  Therefore, the majority of
our sources have X-ray/optical flux ratios that are {\bf larger} than those
displayed.  Figure~\ref{fig7} illustrates the lowest possible X-ray/optical
flux ratios and also, by the sparsity of sources included in the diagram,
it illustrates the fact that a majority of the sources have no obvious
optical counterpart and thus have very large X-ray/optical flux ratios.
We estimate the point source detection limit of the OM U
filter as approximately $1.24 \times
10^{-14}$\,erg\,cm$^{-2}$\,s$^{-1}$.  For an unabsorbed X-ray flux of
$1.0 \times 10^{-12}$\,erg\,cm$^{-2}$\,s$^{-1}$, typical of objects
with L$_X \approx 2 \times 10^{38}$\,erg\,s$^{-1}$ located at a distance
of 8\,Mpc, this corresponds to
$\log(f_{x}/f_{opt}) = 1.9$.  Therefore, the average value for our
sources should fall around 2 or greater.  The average distribution for
QSOs and AGN centers around 0 and 0.8 for BL Lacs \citep{and03}.
Our objects have ratios of L$_x/$L$_{opt}$ at least 10 times
higher than those of AGN and 100 times greater than stars.

Recently, \citet{gut05} identify six ULXs from the catalog of
\citet{col02} as QSOs.  They hypothesize that a large number of ULXs
may in fact be quasars at higher redshift than their supposed host
galaxy.  However, unlike the objects studied in \citet{gut05}, our ULX
sources are all spatially coincident with the optical host galaxy.  In
addition, a majority of our ULXs are not in the proximity of a
noticeable optical point source.  The X-ray/optical flux ratios of our
sources are much larger, on average, than might be expected for a QSO.
It is also worth noting that while some cataloged ULXs may be QSOs,
optical identifications have been made associating other ULXs with a
type B supergiant companion \citep{kun05,liu04}.

\subsection{Classification Criteria}
The spectral fits indicate that to high statistical probability (see appendix A) 
we can distinguish a class of
low-state ULXs from the high-state objects.   This is assuming, as indicated in the introduction,
that ULXs are isotropic emitters with luminosity and spectral form similar to Galactic stellar-mass
X-ray binaries.  In section 3, we stated that our ULX classification depends upon three criteria:
(1) spectral form, (2) luminosity, and (3) location of the X-ray source within the optical host galaxy
(as determined from the DSS images).  We have chosen simple, parametric `non-physical'
models for the spectra because the signal to noise of most of the observations does 
not allow anything else to be constrained.

Of the sources in Table~\ref{tbl-3}, 16 are classified as ``low-state'' objects or low/hard state ULXs, having
unabsorbed luminosities $> 10^{38}$\,erg\,s$^{-1}$ and spectra that are
well fit by power law models.  Throughout this paper, we use the term low-state 
ULXs to include  ``low-state
IMBH candidates'' (sources with L$_X \la 3 \times 10^{39}$\,erg\,s$^{-1}$ and spectra
well-fit by a simple absorbed power law) and
low-state sources with luminosities that clearly classify them as ULX sources
(L$_X \ga 3\times 10^{39}$\,erg\,s$^{-1}$).  These low-state ULX sources are listed in 
Table~\ref{tbl-11}.

In the Spectral Fitting section, we noted that a power law
and high temperature bremsstrahlung model are indistinguishable.  Therefore,
it is important to consider the luminosity of these sources in the claim
that they are not neutron star X-ray binaries accreting at the Eddington luminosity.
Of the low-state ULX sources, only two of the 16 sources have bolometric luminosities
below the Eddington luminosity of a 3\,M$_{\sun}$ object ($\approx 4 \times 10^{38}$\,erg\,s$^{-1}$),
corresponding to the maximum mass of a neutron star.  All of the sources have 
values exceeding the Eddington limit for a 2\,M$_{\sun}$ neutron star.
   
Further, 26 sources have unabsorbed  L$_X \ga 3\times10^{39}$\,erg\,s$^{-1}$,
corresponding to L $\approx$ L$_{Edd}$ at M $> 20$ M$_{\sun}$ as expected
for ``high-state'' IMBHs,
and spectra that are well fit by combined blackbody and power law models.  These
are ``high-state'' objects.  The spectral fits for these sources are listed in
Tables 3 and 4.
In a statistical sense, we find that the greater
the number of counts in the observation the greater our confidence in the thermal 
component contributing to a better fitting model.  We explain our confidence levels
obtained from spectral simulations in the appendix.  

In addition to these high and low
state ULXs, we find a large number of sources best fit by a combined
blackbody and power law model but below our threshold of L$_X \approx
3\times10^{39}$\,erg\,s$^{-1}$ for a high-state ULX (listed in Table 3).  Many of these sources may be
accreting stellar mass black holes with M\,$< 20$\,M$_{\sun}$.  
Some of these non-``ULX'' sources were found
away from the optical extent of the targeted galaxy (from our analysis
of the DSS images), and therefore may be background AGN.

\subsection{Low-State ULXs}
\begin{figure*}
\plotone{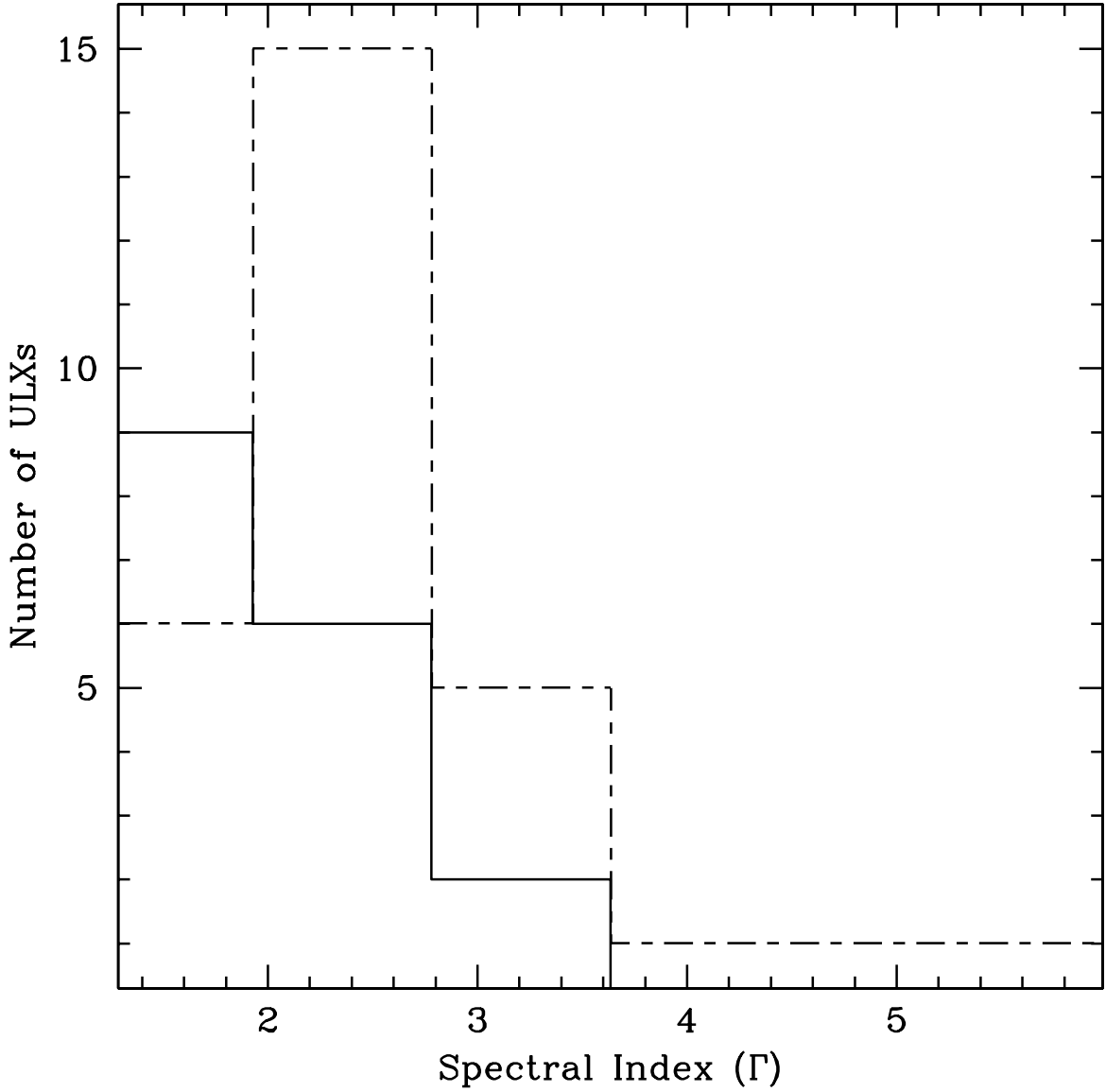}
\caption{Distribution of the spectral indices ($\Gamma$) for low-state
(solid) and high-state (dashed) objects.  For Galactic low-state
objects, typically $\Gamma \approx 2.0$, similar to our sample, while the high-state objects
have a steeper $\Gamma$ \citep{mcc04}. \label{fig2}}
\end{figure*}

For Galactic black hole X-ray binaries, spectral indices of low-state (or power law
dominated) objects are typically lower than those of high-state objects, with
$<\Gamma>_{low} \approx 1.7$ and $<\Gamma>_{high} \approx 2.5$
\citep{mcc04}.  In Figure~\ref{fig2}, we plot the distribution of the
spectral index for both high-state and low-state objects.  The spectral
index for the high-state objects is the value of $\Gamma$ from the
two-component fit.  As in the
Galactic sources, it is clear that the spectral indices of the
high-state objects are indeed larger. Of further interest, the
distribution of spectral index for low-state objects looks remarkably
similar to the distribution of spectral index for moderate luminosity quasars,
many of which are thought to be the analogs of low-state black holes
\citep{por04}.  This supports the classification of these
objects as accreting black holes.

For the high-state objects, we find mean values of $\Gamma = 2.46$, with a root mean square (rms)
deviation of S$ = 0.12$, and L$_{X} = 1.4\times10^{40}$\,erg\,s$^{-1}$, $\log(S) = 1.6$.  
This calculation excludes the 3 objects with spectral
indices greater than 3.5.  For the low-state objects, we find mean values of
$\Gamma = 2.09$, with a rms deviation of S$ = 0.10$, and L$_{X} = 2.2\times10^{39}$\,erg\,s$^{-1}$,
$\log(S) = 2.1$.  This value of $\Gamma = 2.1$ is softer than the typical hard-state value of
$\approx 1.7$, but within the $1.5 < \Gamma < 2.1$ range used to classify this state for
Galactic X-ray binaries \citep{mcc04}.
Computing a Kolmogorov-Smirnov two-sample test, separating the sources
into the category of low-state or high-state, we find a likelihood of approximately 0.03
that the spectral indices belong to the same distribution. 

\begin{figure*}
\plotone{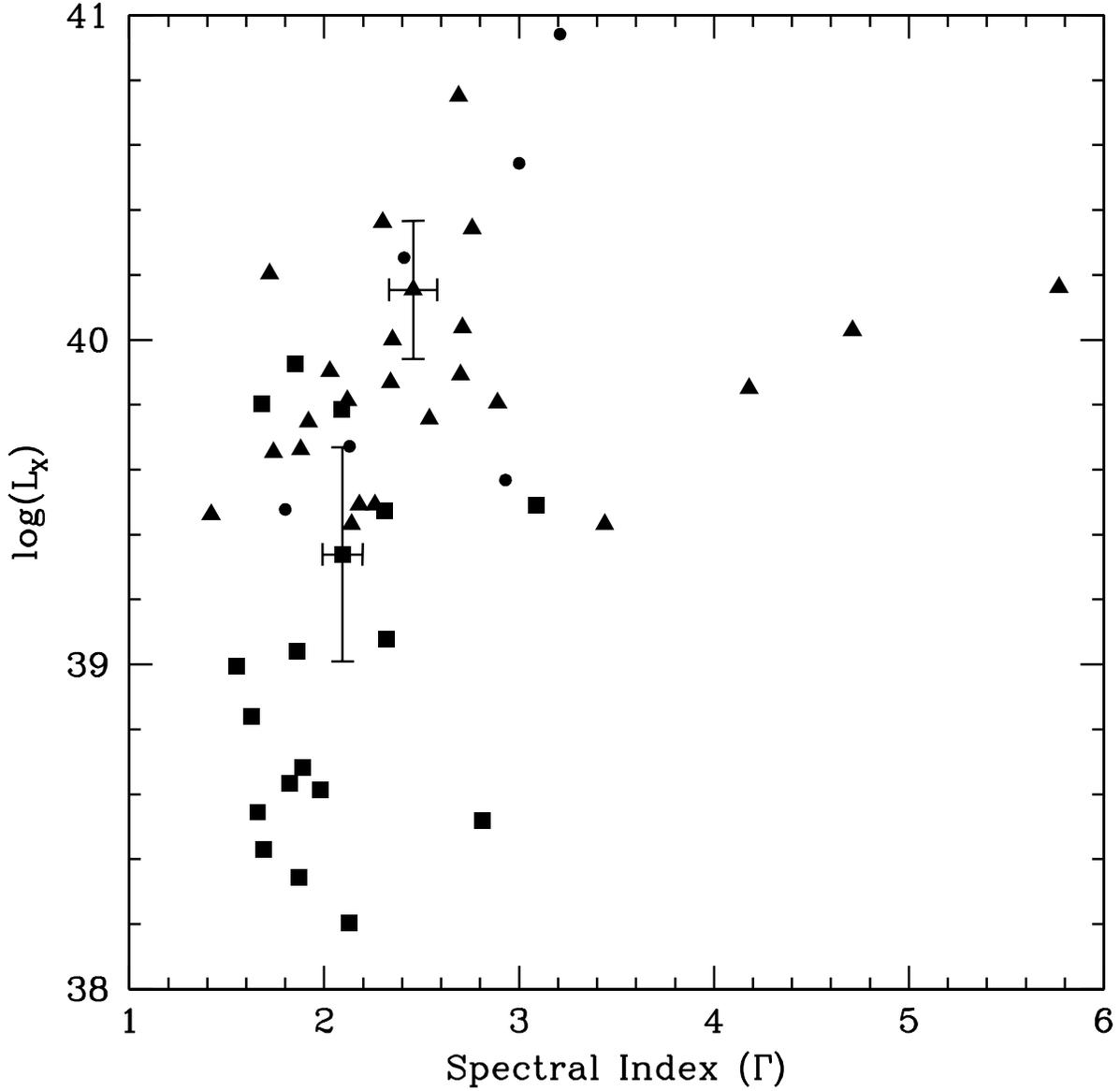}
\caption{Relationship of luminosity vs. spectral index for low-state
(rectangle) and high-state (triangle) objects. Sources represented
by a circle are those where the $\Delta \chi^2$ value between the
two-component and power law fits was very small.
As expected from
observations of Galactic stellar-mass black hole systems
\citep{mcc04}, the classified low-state ULXs in our sample have, on
average, lower X-ray luminosities than the corresponding high-state
ULXs. We plot the mean values for both high-state and low-state objects
with errorbars indicating the root mean square deviation.  The outlying
objects with spectral indices greater than 3.5 were not included in
the mean or deviation calculations.\label{fig3}}
\end{figure*}

The low hard X-ray state of X-ray binaries is associated with a low
accretion rate from the companion object with
L$\la 0.1$\,L$_{Edd}$ \citep{don03}.  Therefore, on average, we
expect the luminosities of the low-state objects to be lower than the
high-state objects.  Figure~\ref{fig3} displays the luminosity of the objects as a
function of the spectral index.  On average, the highest luminosity
low-state objects have luminosities lower than those of the high-state
objects.  

The lower L$_{X}$ values of the low-state objects imply that they may
indeed be accreting at a lower rate than the high-state objects.  This can
further be seen in the bolometric luminosities listed in Table ~\ref{tbl-11}.
If these objects are accreting at a rate similar to Galactic low/hard state
black holes ($0.1 \times L_{Edd}$) \citep{don03}, we can estimate their masses 
as 
\begin{displaymath}
\frac{M}{M_{\sun}}= \frac{L_{bol}}{0.1 \times L_{Edd}}
\end{displaymath}
with $L_{Edd}$ as the Eddington luminosity for a 1\,M$_{\sun}$ object ($1.3 \times 10^{38}$\,erg\,s$^{-1}$).
Our mass estimations, based upon our limits to the bolometric luminosities,
yield masses of $20 - 1500 M_{\sun}$ (see Table ~\ref{tbl-11}), precisely what we might expect for a 
population of IMBHs.

\subsection{High-State ULXs}
\begin{figure*}
\plottwo{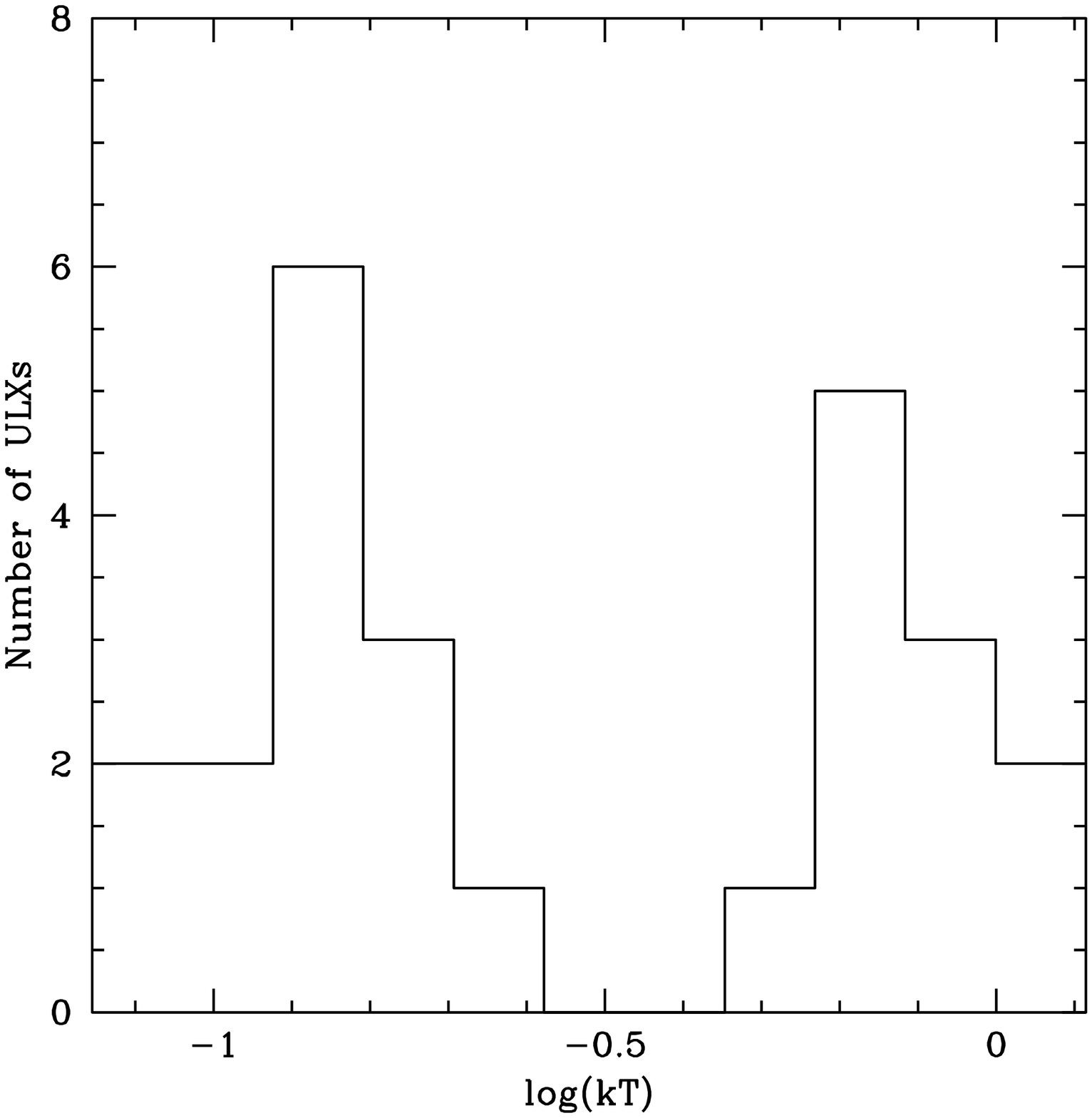}{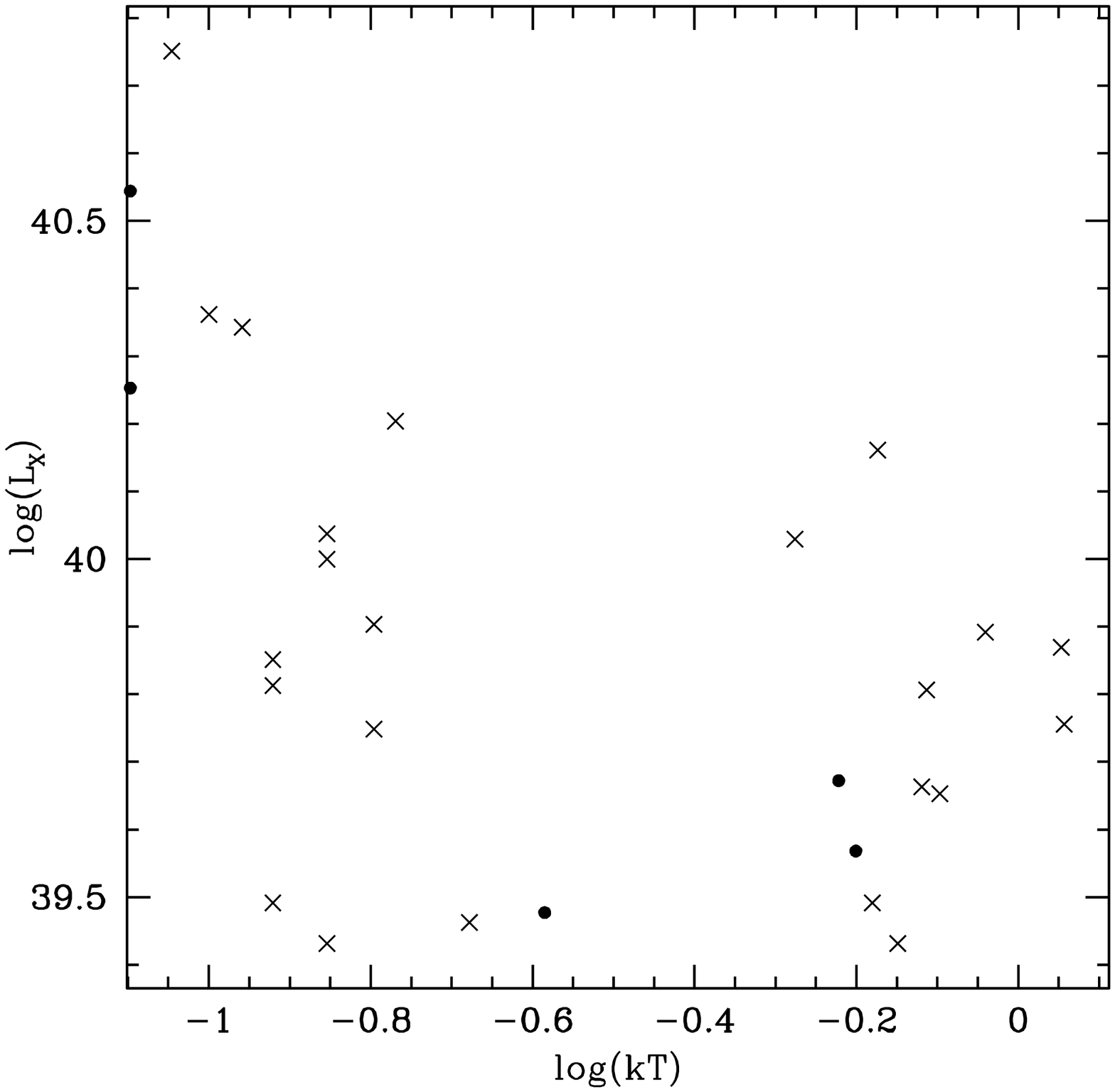}
\caption{(left)Distribution of the blackbody temperature for
high-state objects.(right)Relationship of blackbody temperature
vs. luminosity (in the 0.3-10 keV band) for high-state objects. We see
two peaks arise in the distribution, one centered around kT$\approx 0.1$
and another at kT$\approx 1$.  The peak with a low disk
temperature also corresponds to the highest luminosities, suggesting
that these may be high-state IMBHs.  The sources with higher disk
temperature also have lower luminosities.  The spectra of these
sources were also well fit by an inverse Comptonization model (a model
succesfully used to fit some of the Galactic black hole X-ray binaries in the
very high state). \label{fig4}}
\end{figure*}

If the high-state (thermal dominated) ULXs represent a class of intermediate mass black
hole systems, their X-ray spectra should be well described by a combined
blackbody and power law model.  Scaling for the mass of the black hole,
we would expect a relationship of T\,$\propto$\,M$^{-1/4}$ between
black hole mass and blackbody temperature \citep{mak00}.  This would
indicate a thermal component of $\sim$ 100\,eV for masses of $\sim 10^3$\,M$_{\sun}$.  A few objects have
been reported to display this property \citep{mil03,rob05}.  In
Figure~\ref{fig4}, we graph the distribution of the thermal component
for our classified high-state objects.

We find that there are two peaks in the distribution among the thermal component,
one at approximately 100 eV and another centered close to 1 keV.  This
could indicate two different classes among the high-state objects.  It
is possible that those objects with blackbody components near 100\,eV
are indeed high-state intermediate mass black holes.  We note that the soft
excess in PG Quasars has also been modeled as a blackbody with kT$_{soft} \approx 100$\,eV,
but it has been suggested that this could be the result of a process not directly related
to black hole accretion (such as the presence of a warm absorber: \citet{gie04}).
Another possible explanation is that the soft component is the result of ionized reflection
from the disk \citep{ros05}.
While the possibility exists that the ``thermal'' component of these 100\,eV sources
is not directly related to black hole accretion or is related in a ``non-thermal'' (i.e. ionized reflection)
sense, as may be the case with the soft excess in PG Quasars,
we assume that the soft component for the objects we classify as high-state ULXs originates
from a thermal disk.  We use this assumption to test the IMBH hypothesis, thus
speculation on the nature of the soft component is beyond the scope of this paper.  

The second peak, centered around 1 keV, has a temperature
reminiscent of the Galactic black hole systems in our own galaxy.
These systems may thus be stellar-mass black holes accreting matter near the
Eddington limit.  If this were the case, we would expect the luminosities
of the sources exhibiting a higher blackbody temperature to be lower than those
with cooler blackbody components.
In the second graph of Figure~\ref{fig4}, we plot the relationship between blackbody temperature
and L$_X$ in the 0.3 - 10 keV band.  Once again, two groups are seen in the distribution 
of high-state ULXs.  The most luminous objects are those with low blackbody
temperatures.  On average, the less luminous sources exhibit higher
blackbody temperatures.  For the sources with L$_X > 10^{40}$\,erg\,s$^{-1}$,
the mean blackbody temperature is 0.31 while the sources below this luminosity
threshold have a mean blackbody temperature of 0.61.

The second, low-luminosity, group in the distribution of high-state ULXs 
is clearly distinguishable in
both plots of Figure~\ref{fig4}.  We found that, with the exceptions of NGC 253
XMM1, M81 XMM1, and NGC 5204 XMM1, the spectra of these objects could also
be well-described by an absorbed Comptonization ({\it compST}) model
\citep{sun80} used to fit galactic black holes in the ``very high''
state when they are radiating at the Eddington limit.  This model
simulates Compton scattering of cool photons on the hot electrons of a
completely ionized plasma.  We present the best-fit parameters for the
Comptonization model in Table~\ref{tbl-9}.

This ``very high'' state has been observed \citep{miy91} in a few
Galactic black holes.  Yet another rubric for the very
high state emerged in \citet{kub01} and \citet{kub04}, where they
identify this as the ``anomalous'' state, a state whose spectrum can
be well fit by a Comptonized scattering model.  Regardless of the
name, our best-fit Comptonization sources likely fit into this
category.  The luminosities of these sources suggest that they are
stellar mass black hole systems in this anomalous/very high state.

As with the low-state, we include mass estimates for our
high-state objects in Table~\ref{tbl-11}.  We assume that the
high-state objects are radiating at L$_{Edd}$ resulting in a minimum
mass if there is no beaming.  We find masses of
$1.6 - 38$\,M$_{\sun}$, consistent with ``normal'' stellar mass BHs,
for the sources well fit by the Comptonization model.  The other
high-state ULXs masses range from $17 - 1350$\,M$_{\sun}$ based on Eddington
rates, analogous to
the low-state ULX masses computed.

It is important to note that the initial simulations (appendix) and
discussions in the Spectral Simulations section need to be considered
in relation to the impact they pose to our classification scheme
and the results presented in these sections.    While it is indeed possible 
that some of the objects with a weak
blackbody component and a relatively small number of counts would be
mis-categorized as a pure power law spectrum, one can ask what such a
possible situation would do to the correlations that we have seen.
These putative objects, by assumption would have lower luminosities,
however their temperatures are unknown and it is entirely unclear if
they would destroy the kT/L(x) correlation. As we have shown in our
simulations it is unlikely that the fitted power law index would
change and thus the presence of a low state as indicated by the
spectral index would not change. This would create a new type of
object, one with a flat power law and a black body component, which is
not seen in the Milky Way, nor among the high signal to noise
objects.

\subsection{Temperature Gap}

In addition to the existence of ULXs with low blackbody temperatures,
the temperature distribution of the high-state (thermal dominated)
ULXs (Figure~\ref{fig4}, left panel) displays a ``gap''
which is of particular interest --- there is a complete absence of
objects with temperatures in the range 0.26\,keV to 0.50\,keV.  It is
tempting to take this as evidence for a gap in the mass distribution
of these accreting black holes.  Since, for a given luminosity, we
expect the temperature to vary as $T\propto L^{1/4}M^{-1/2}$, this
factor of two gap in the temperature distribution translates into a
factor of four gap in the black hole mass distribution.

If this result is borne out by further study, it provides an important
clue to the origin and evolution of intermediate mass black holes.
One popular idea is that intermediate mass black holes formed from the
collapse of massive Population III stars \citep{mad01}.  Models
suggest that Pop III stars with zero age main sequence (ZAMS) masses
in the range 25--140$M_\odot$ and above 260$M_\odot$ collapse to
produce black holes \citep{heg02} whereas in the range of
ZAMS masses 140--260$M_\odot$, pair-instability supernovae lead to the
complete disruption of the stars (i.e., no remnant black hole
remains).  Hence, this model for IMBH formation predicts a gap in the
IMBH initial mass function in the range of approximately
60--200$M_\odot$ (although this is uncertain on the low end due to the
effect of the pulsational pair-instability on the pre-collapse core).
One possibility is that the gap in our observed temperature
distribution (and hence the inferred gap in the mass function) is due
to this effect of the pair instability supernovae in Pop III stars.
This would require that the current IMBH mass function is
approximately the same as the initial IMBH mass function.  In other
words, it requires that most IMBHs (especially those just below the
gap) have not grown significantly due to accretion since their
formation and, hence, that the ULX phase represents a short fraction
of the life-time of an IMBH ($f<<t_{\rm sal}/t_{\rm H}$, where $t_{\rm
sal}\approx 45\epsilon_{0.1}\,Myr$ is the e-folding timescale for
Eddington limited black hole growth with radiative efficiency
$\epsilon=0.1\epsilon_{0.1}$).

An alternative interpretation of the inferred mass gap is to suppose
that two fundamentally different modes of formation lead to a strong
bi-modality in the final black hole mass function.  Black hole masses
below the gap can be readily understood through normal stellar
processes.  A separate and distinct population of significantly more
massive black holes may result from dynamical processes in the core of
dense globular clusters \citep{mil02,gul04}.

\subsection{Comparison with Galactic HMXBs}
Supposing that the Galaxy's bright X-ray population is representative
of low-redshift galaxies, we expected to find a number of sources
similar to Galactic X-ray binaries in our sample.  In our
sample, we find approximately 24 sources with luminosities below our
high-state ULX cutoff (${\approx} 3 \times
10^{39}$\,erg\,s$^{-1}$) , X-ray positions within the optical extent of
their host galaxy, and no obvious optical counterpart.  The unabsorbed
luminosities for these sources range from $0.4-2.5 \times
10^{39}$\,erg\,s$^{-1}$ ($0.3 - 10$\,keV band).  Two of
these sources were transients in the {\it XMM} data.  Of the four host galaxies with
multiple observations examined, two of these galaxies contained solely
ULX sources in our luminosity regime (Holmberg II and NGC 5204).  Each
of the remaining two (NGC 253 and NGC 4258) had a transient source
best fit by a combined blackbody and a power law.

This suggests an interesting diagnostic in terms of distinguishing
our ULX sources from a normal HMXB population.  In our own galaxy,
most HMXBs vary on timescales of days or less and most of the black
holes in the Milky Way are transients, though some HMXBs are indeed
persistent.  The figures in 
\citet{kal04}, determined through detailed mass-transfer calculations, 
indicate that transient behavior should not be expected
from a population of IMBHs.
Thus, on average, our ULX sources should remain X-ray bright in
multiple observations.  Through a literature search, we found that
37/42 of our ULX sources are well detected in ROSAT observations and thus
are luminous for greater than 10 years and therefore are not
transients.  Examination of the long term light curves show that most
of these sources vary by less than a factor of 3 over the timescale from
ROSAT to XMM.  The sources that have been above the
Eddington limit in the Milky Way and the Magellenic clouds do so
transiently, for a small fraction of the time.  As best as we can tell, 
from the light curves from Einstein, ROSAT, ASCA, Chandra and XMM
the ULXs are, rarely, transients, and are almost
always `on', unlike Galactic ``ultra-luminous'' objects.

\begin{figure*}
\plotone{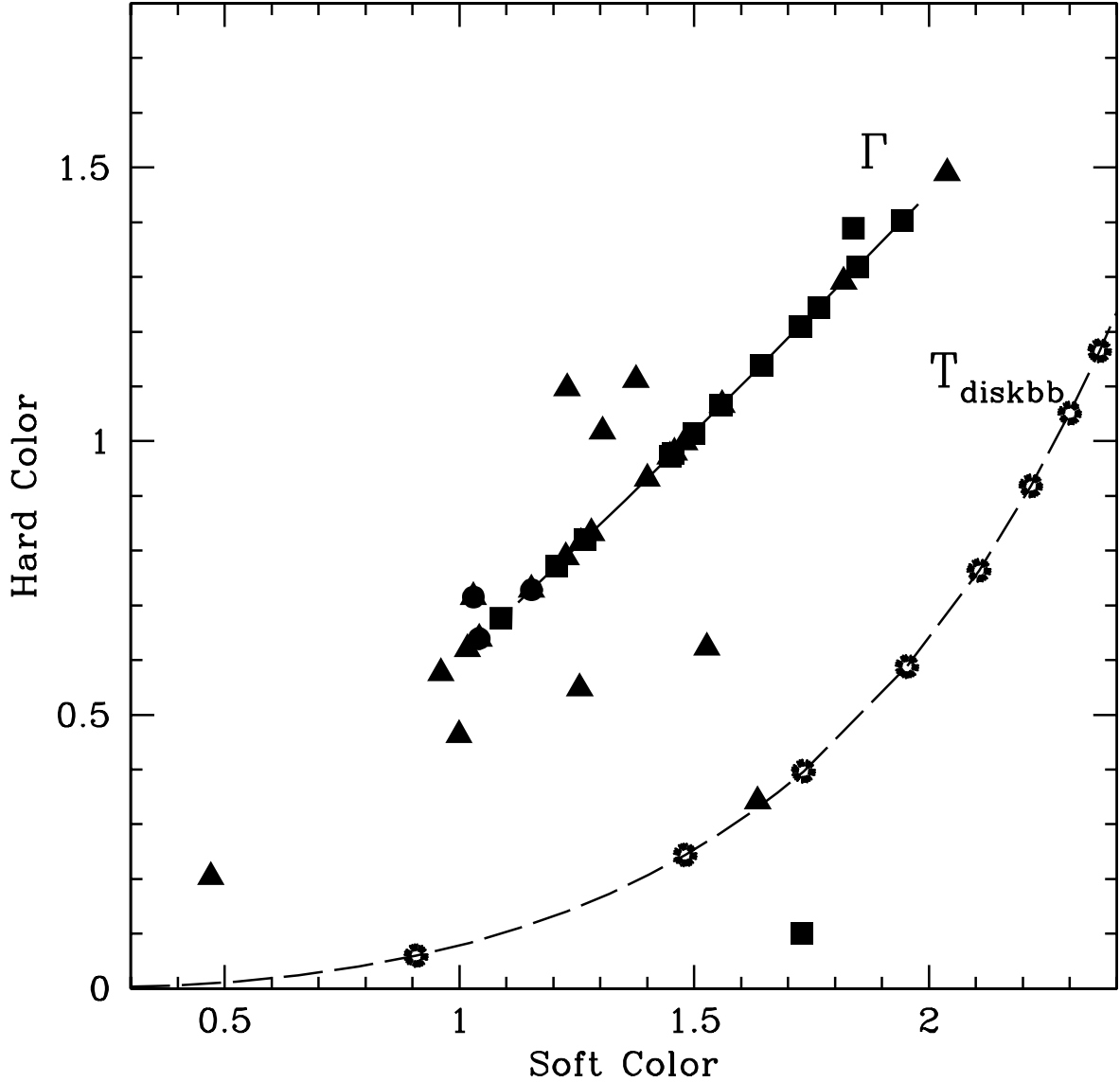}
\caption{Color-Color Diagram plotting soft vs. hard colors, as
outlined in \citet{don03}, for low-state (rectangle) and high-state
(triangle) ULXs.  The filled circles represent the sources with
low $\Delta\chi^2$ values between the two-component and power law models.
A large number of our sources lie in the same range
of this graph as the black hole sources examined by \citet{don03}
(near the power law distribution, indicated by the solid line).  The
dashed line represents the color-color plot for a multi-colored disk
model with different disk temperatures.  The sources approaching this line were those
well-fit by the Comptonization model.  \citet{don03} had no black hole
sources in this region, but atolls and Z-sources, which were also
well-fit by Comptonization models.
\label{fig5}}
\end{figure*}

As a possible further diagnostic, we constructed a color-color diagram
for our ULX sources.  We adopted the colors of \citet{don03} in order
to compare our sample with their sample of Galactic X-ray sources.
Thus, our colors were constructed from unabsorbed model fluxes in four
energy bands: 3-4, 4-6.4, 6.4-9.7, and 9.7-16 keV.  The {\it XSPEC}
command {\it dummyresp} was used to calculate a flux based on the
model for the 10-16 keV range.  We plot colors for a pure unabsorbed
power law (from $\Gamma = 1.5-3.0$) and an unabsorbed MCD model ({\it
diskbb} in {\it XSPEC} with kT$_{in} = 5.0-0.2$\,eV) for comparison.
Comparing our Figure~\ref{fig5} with Figure 8 of \citet{don03}, we
find that our ULX sources largely lie along the same regions as their
black hole sources.  A few ULX sources, however, lie in the region
occupied by atoll and Z-sources in the plot of \citet{don03}.  These
sources were those best fit by a Comptonization model.

\subsection{Galaxy Sample}
 In this section, we examine the environment in which ULX sources
reside.  We investigate the claim that
the ULX population is proportional to the
host galaxy's star formation rate (SFR) (Ranalli et al. 2002; Grimm et
al. 2003).   Towards this end, we use the far-infrared luminosity of the host galaxy as an
indicator of the SFR.  In order to compare the ULX population of a
galaxy with the SFR we followed a similar approach to \citet{swa04}.
We calculate the FIR flux from observations taken by the {\it Infrared
Astronomical Satellite}.  As in \citet{swa04}, the flux between 42.4
and 122.5 $\mu$m is approximated as 1.26$\times 10^{-11} (2.58$S$_{60}
+$S$_{100})$\,erg\,cm$^{-2}$\,s$^{-1}$.  The values of the flux at
60\,$\mu$m (S$_{60}$) and 100\,$\mu$m (S$_{100}$) were obtained from
either \citet{ho97} or NED.  Luminosities were calculated using the
distances quoted in Table~\ref{tbl-1}.  We list these values in
addition to the number of ULXs observed in individual galaxies in Table~\ref{tbl-10}. 
The number of ULXs includes both the objects we
classify as high and low state ULX as well as those sources resolved
by Chandra.

\begin{figure*}
\plottwo{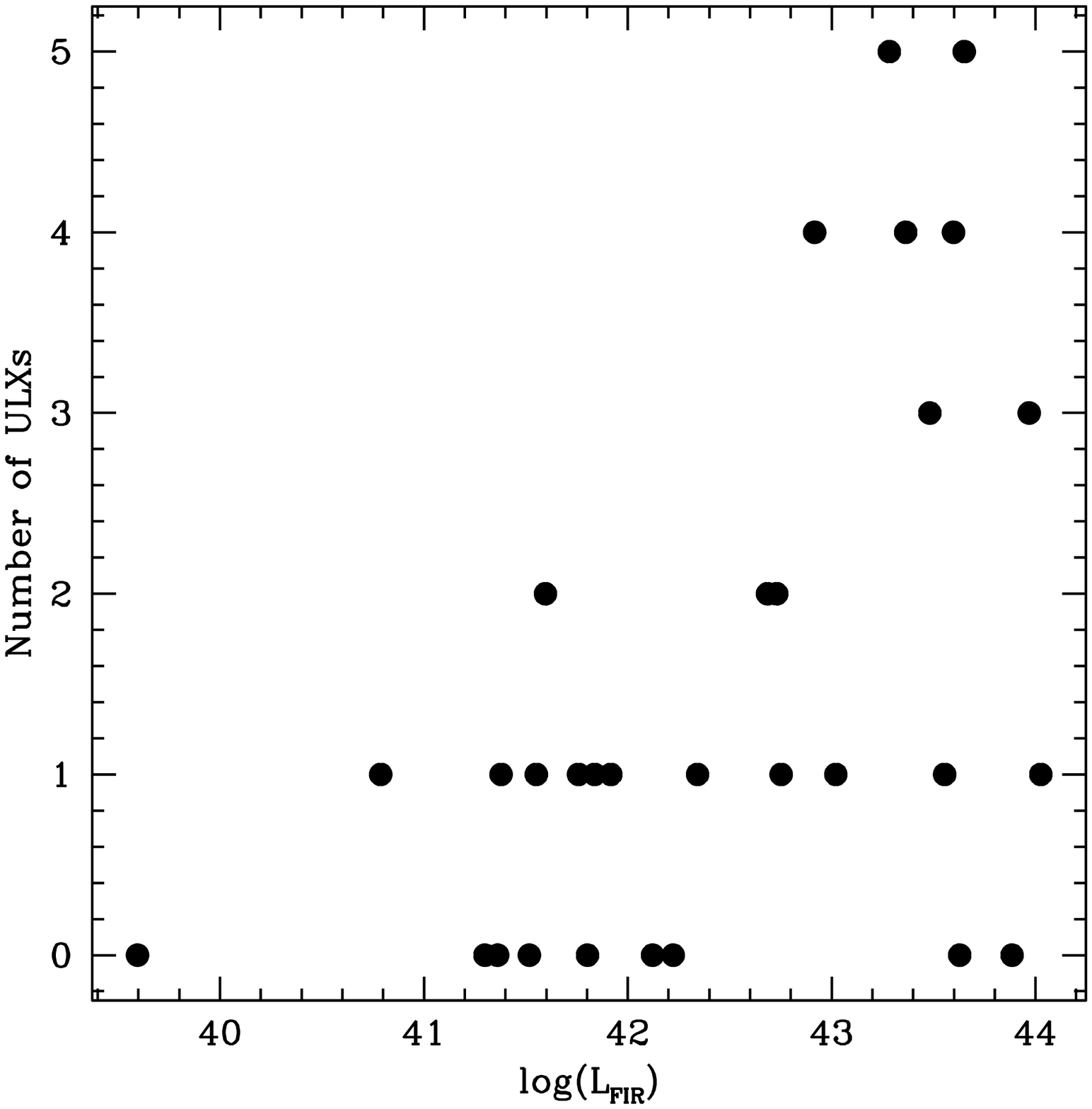}{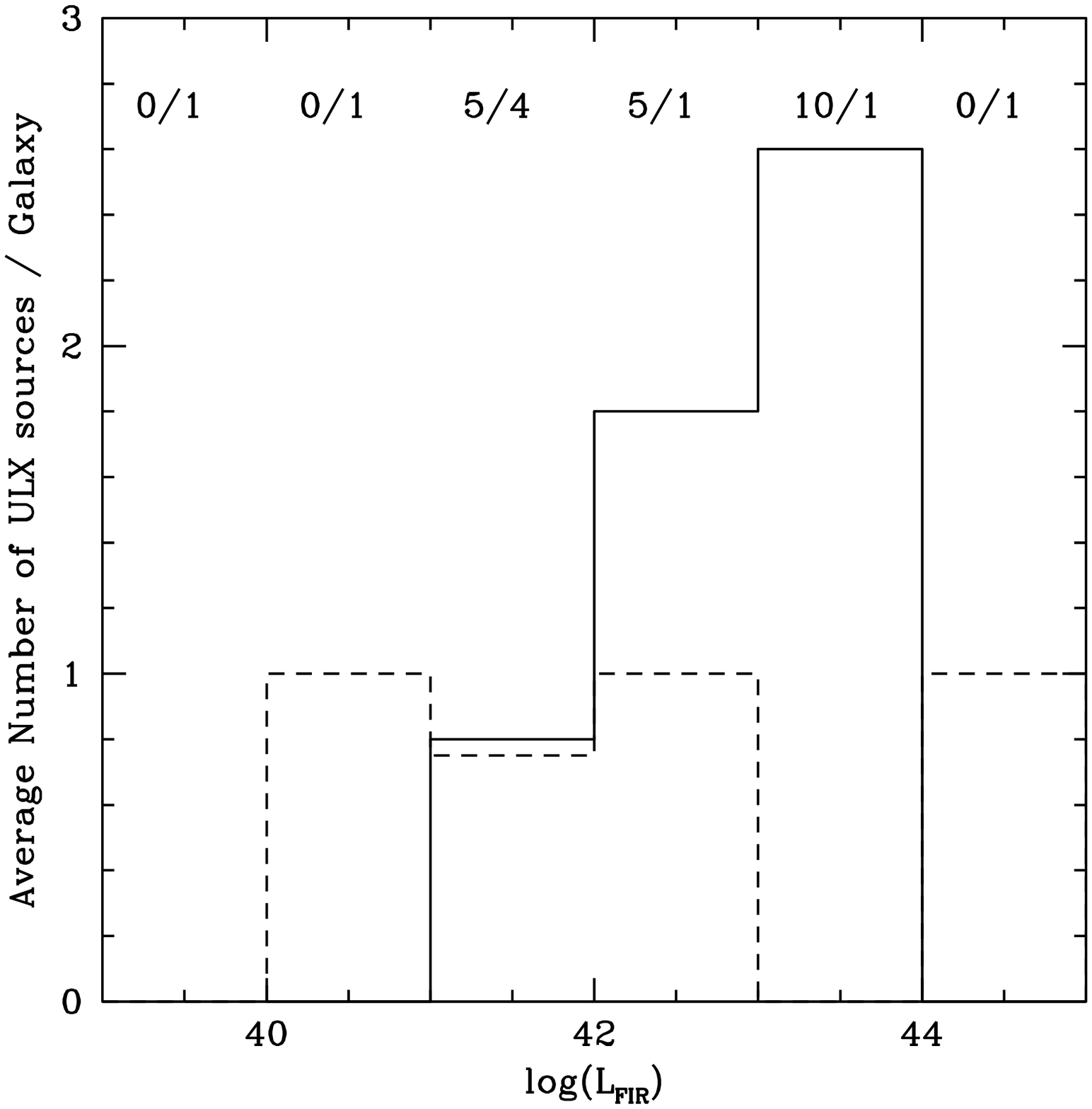}
\caption{(left) Relationship of the far-infrared luminosity, as an indicator
of star formation rate, vs. the number of ULXs for each galaxy.
If ULXs are associated with star formation, we naively expect that the
higher the FIR luminosity the more ULXs the galaxy will host.  (right)
The distribution of average number of ULXs / L$_{FIR}$ bin for spirals
(solid line) follows this expectation.  The distribution of irregulars
(dashed line) is not so easily interpreted.  The numbers at the top
indicate the number of spirals/irregulars in each of the luminosity
bins.  More irregulars would need to be included in this survey for
meaningful statistics on this group.
\label{fig6}}
\end{figure*}

In Figure~\ref{fig6}, we show two plots relating the number of ULXs
to L$_{FIR}$.  It has been suggested by \citet{gri02} that the luminosity 
function in the X-ray regime from HMXBs is related to SFR.  In our first plot,
we find that the galaxies with the highest L$_{FIR}$ seem to have fewer
ULXs than may be expected from the luminosity functions of \citet{gri02},
who present a relationship showing a scaling of the number of HMXBs with
luminosities over a set threshold with the host galaxy's SFR 
(see equation 7; Grimm et al. 2002).  Using this relationship, we would
expect that a galaxy with a SFR approximately equal to that of M51 ($\approx 4$\,M$_{\sun}$\,yr$^{-1}$ from their table 1) 
to have $\approx 4.47$ objects with luminosities greater than $10^{39}$\,erg\,s$^{-1}$.
We find 5 objects with this luminosity in M51, consistent with their result.  However,
for NGC 4945, a galaxy with approximately the same L$_{FIR}$ and therefore SFR, we
find only one source with a luminosity in this range. However, we note that
NGC 4945 is a Seyfert, implying that the L$_{FIR}$ may primarily be caused by the AGN and not a direct
indication of SFR. In addition to high L$_{FIR}$ sources with few ULX we find a number of
sources with very small SFR but which contain a ULX.  For sources with SFR $< 0.2$\,M$_{\sun}$\,yr$^{-1}$,
which corresponds roughly to sources with L$_{FIR}$ less than that of NGC 4736,
we would expect $< 0.22$ sources with luminosities above $10^{39}$\,erg\,s$^{-1}$.
However, there are a number of bright ULXs in galaxies with very low SFRs (for instance
Holmberg II, Holmberg IX, NGC 5204, and NGC 5408).    
Thus, in a direct comparison, our results do not agree with the predictions of \citet{gri02}. 

The second plot displays the average number of
ULXs/galaxy, binned according to luminosity.  This plot is extremely
similar to Fig. 15 of \citet{swa04} for spiral galaxies.  Thus, once
again, it seems that the connection between SFR and the ULX population
in spirals is supported.  For irregular galaxies, however, there seems
to be more of a spread in the distribution.  This could be the result
of poor sampling --- most of the bins contain only one galaxy.  Another
possibility is that there is no direct correlation in irregular
galaxies or that the overall star formation in these galaxies is less
ordered or clumpier.  If the latter is the case, the overall SFR of
the galaxy is only an average over a wide range of values.  We shall
address this issue again in the next paper in this series (L.M. Winter et
al., submitted) where we discuss the local environments of the ULXs in
our sample.

\begin{figure*}
\plotone{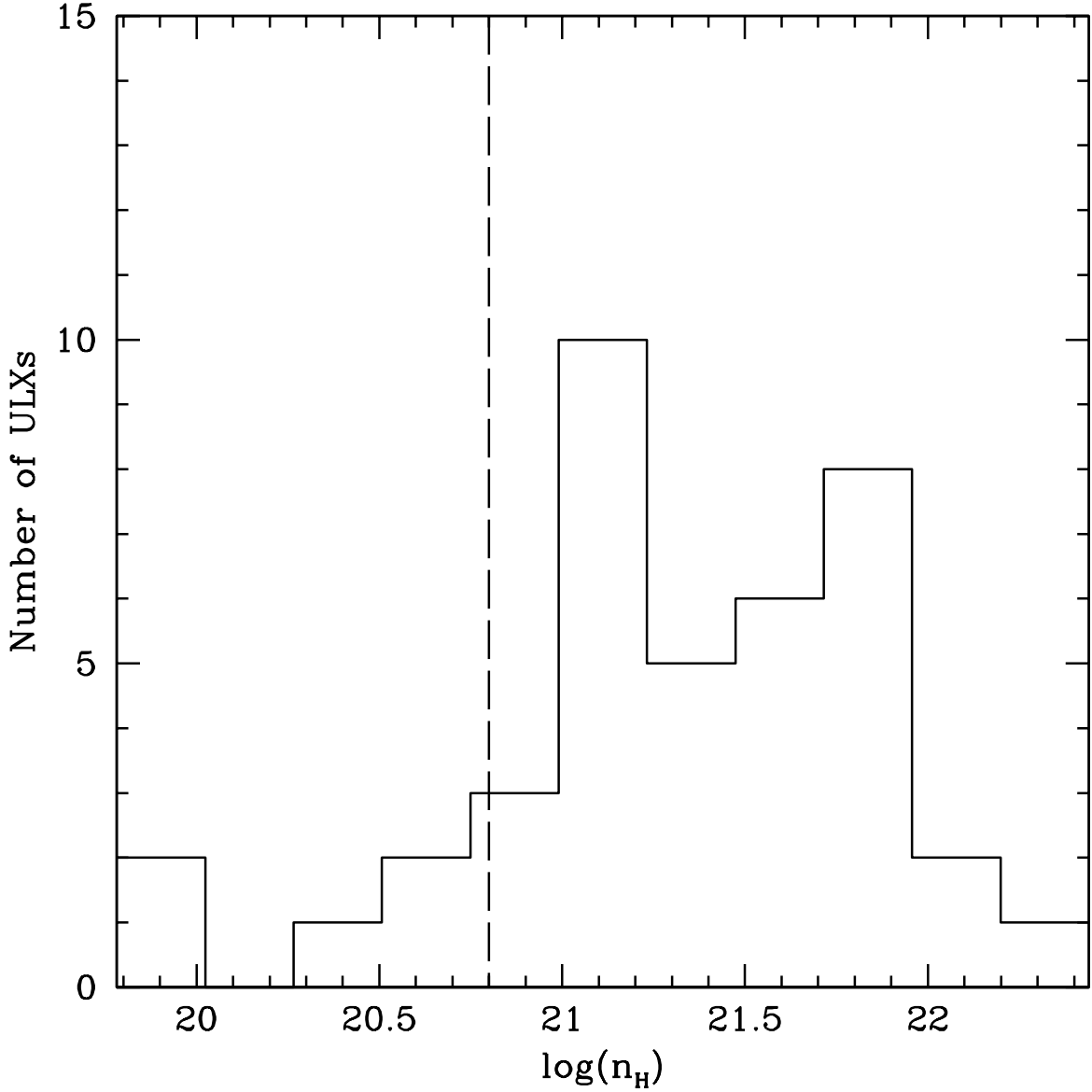}
\caption{Distribution of the hydrogen column densities of ULX sources. The
n$_{H}$ values were obtained through spectral fits using the {\it wabs} model
in {\it XSPEC}.  Galactic column densities towards the host galaxy were
subtracted from the spectral fit values. A majority of our ULX sources have
high column densities ($ > 10^{21} cm^{-2}$), suggesting that some of this
absorption originates with the local ULX environment. Bins to the left of the dashed line represent
sources with column densities very close to the Galactic value and thus a simple
subtraction is not statistically representative of the true value.
\label{fig8}}
\end{figure*}

In Figure~\ref{fig8} we plot the distribution of column densities among
the ULXs.  We subtracted the Galactic column density towards the galaxy
(obtained from the nH FTOOL and listed in Table~\ref{tbl-1}) from the
values obtained through spectral fits.  We note that, on average, the
ULXs have large column densities.  The typical Galactic column density
along a line of sight is $\approx 4\times 10^{20}$\,cm$^{-2}$.  If the
ULX is located on the opposite side of its host galaxy, we might expect
maximum column densities of $\approx 1.2\times 10^{21}$\,cm$^{-2}$.
However, most of our sources have column densities well above this value.
This is in agreement with the analysis of 5 ULXs by \citet{rob04} and
may imply, as they suggest, that the local environment of the ULXs contains an extra
source of absorption.  We are investigating this further, comparing the
X-ray absorption column densities with HI data (L.M. Winter et al., submitted).

In order to better understand the relationship between SFR and the ULX
population, it is necessary to extend ULX studies to other
wavelengths.  In particular, it becomes important to analyze UV and IR
images close to the ULX.

\section{Conclusion}

 We present the results of an {\it XMM} survey of the ULX
population in nearby galaxies.  In this study, we assumed that ULXs are isotropic emitters.  
For our selected ULX sources (which excluded transient sources and supernovae), this 
assumption was supported by the finding that 37/42 of our ULXs were 
found to be `on' in ROSAT
observations.  This implies that theses sources exhibited high luminosities for time scales of
at least 10\,years, a property that is not seen in Galactic Eddington-limit exceeding sources (such
as black hole X-ray binaries undergoing an outburst).  
We also assumed that if some ULX sources represent a class of
IMBH X-ray binaries, they would exhibit spectral states analogous to Galactic stellar mass black hole 
X-ray binaries.  This is the hypothesis we set out to test, classifying a source as a ULX based on
(1) spectral form, (2) luminosity, and (3) coincidence of the X-ray source within the optical host galaxy.
Due to the quality of spectra available for these distant X-ray sources, our classification of spectral
form is really a first approximation describing the basic curvature of the spectrum. 

 Through this study, we have found that there exists a population of objects whose X-ray spectral 
properties closely match the low/hard state spectra of Galactic black holes, 
but whose luminosities lie in the range of L$_{bol} \approx 2 \times 
10^{38} - 1 \times 10^{40}$\,erg\,s$^{-1}$.  In the Milky Way, black holes 
with these spectral properties radiate at only $\approx 0.05$ of the Eddington limit.  If this is also 
true for this population, it indirectly implies that these objects have a mass greater than 
$\approx 30$\,M$_{\sun}$ ranging up to 1500\,M$_{\sun}$ and thus should be IMBHs.  The existence of 
such objects was ``predicted'' on the basis that the ULXs previously studied shared the X-ray spectral 
characteristics of high-state Galactic black holes; namely, an X-ray spectrum best fit by a combined 
blackbody and a power law \citep{mil03}, but with much higher luminosities.  If these objects are high-state 
IMBHs, the corresponding low-state objects should also exist.  

Our survey has also uncovered a large population of objects whose X-ray spectra are well modeled by the 
canonical description of Galactic black holes in the high-state (thermal dominated), a black hole with a steep power law, but 
whose bolometric luminosities exceed $2 \times 10^{39}$\,erg\,s$^{-1}$, ranging up to $10^{41.5}$\,erg\,s$^{-1}$ 
and whose blackbody temperatures are less than 0.3\,keV.  
If these objects are radiating at $\approx 1/2$ the 
Eddington limit like their Milky Way counterparts their implied masses are from $30 - 3000$\,M$_{\sun}$, 
a range very similar to that implied by the low-state objects.  Using the M$^{-1/4}$ scaling of mass to 
temperature, the observed spectral temperatures give masses of $500 - 10^{4}$\,M$_{\sun}$ a considerably 
larger value.  In general agreement with the expectations of the IMBH hypothesis, the objects with high-state 
spectra are more luminous than those with low-state spectra.  We note 
that these results have required the high signal to noise of {\it XMM} in order to discern the spectrum of 
these objects.  Many of these objects have also been observed by Chandra and their spectra have been 
well-fitted by simple power laws.

In addition to classification of the sources, we investigated some of the properties of the
ULX sources.  We found a gap in the temperature distribution of high/soft state ULXs.  This
gap may indicate a gap in mass distribution, which may provide clues to the nature of ULXs.
We also found that our ULXs are persistent sources (not transients) which occupy regions
on the color-color diagram of \citet{don03} also occupied by Galactic black hole sources. 
Lastly, the existence of a substantial population of ULXs in nearby dwarf and other low star formation rate galaxies 
argues that (in agreement with \citet{pta04,swa04}) there is more than one source term for the origin of ULXs,
 with at least some of them not being associated with recent star formation, at least statistically.

We conclude, from an X-ray spectral and luminosity point of view, that our data are consistent with many of 
these objects having the properties expected of an IMBH population.  However, we also find two other 
populations of objects, those whose blackbody temperature and luminosity correspond to that of stellar 
mass black holes with kT $\approx 1$\,keV and $\log L_X$ less than $2 \times 10^{39}$\,erg\,s$^{-1}$ and 
a small population of objects whose X-ray spectra and luminosities are consistent with that of stellar 
mass black holes in the very high state.  Thus, ULX selected purely on the basis of $0.3 - 10$\,keV X-ray 
luminosities are a composite class with $\approx 1/4$ being ``normal'' stellar mass black holes and the 
rest being consistent with a population of IMBHs.

In a follow-up paper we will discuss the environments of these objects
as revealed by {\it XMM} OM UV imaging and the implications this has
for the origin of ULXs.

\acknowledgements
L.W. gratefully acknowledges Kip Kuntz and M. Coleman Miller for helpful discussions. 
We would also like to acknowledge R. Narayan for asking the question ``are there any low-state
ULXs?'' at the Kyoto Black Hole meeting.



\clearpage
\begin{deluxetable}{llllllll}
\tabletypesize{\scriptsize}
\tablecaption{{\it XMM-Newton} Galaxy Observations\label{tbl-1}}
\tablewidth{0pt}
\tablehead{
\colhead{Galaxy} & \colhead{Type\tablenotemark{a}} & \colhead{n$_H$\tablenotemark{b}} & \colhead{distance\tablenotemark{c}} &
\colhead{ref} & \colhead{obs id\tablenotemark{d}} & \colhead{duration (s)} & \colhead{comments}
}

\startdata
NGC247      & SAB(s)d         & 1.54 & 3.09& \nodata & 0110990301 & 14536 & - \\
NGC253      & SAB(s)c;HII     & 1.40 & 3.73& \nodata &  0110900101, 0152020101 & 30711, 110591 & Starburst \\
NGC300      & SA(s)d          & 3.11 & 2.56& \nodata & 0112800101 & 43967 & - \\
NGC625      & SB(s)m? sp; HII & 2.15 & 2.62& \nodata & 0085100101 & 26288 & - \\
NGC1313     & SB(s)d; HII     & 4.0  & 4.17& \nodata & 0106860101 & 41310 & - \\
IC0342      & SAB(rs)cd; HII  & 30.3 & 3.9 & 1       & 0093640901 & 11217 & - \\
NGC1569     & IBm	      & 21.7 & 1.6 & 1       & 0112290801 & 15582 & Starburst \\
NGC1705     & SA0- pec; HII   & 3.9  & 5.1 & 2       & 0148650101 & 58926 & Starburst \\
MRK 71      & BCD; HII	      & 3.9 & 3.4  & 3       & 0141150201 & 45919 & galaxy pair \\
NGC2403     & SAB(s)cd; HII   & 4.15 & 3.56& \nodata & 0150651201 & 11415 & - \\
Holmberg II & Im              & 3.42 & 2.70& \nodata & 0112520701, 0112520901 & 13528, 6860 & - \\
Holmberg I  & IAB(s)m         & 3.49 & 3.6 & 4       & 0026340101 & 26280 & - \\ 
M81         & SA(s)ab;LINER   & 4.12 & 3.6 & 4       & 0111800101 & 127913 & Hol IX also in field of view \\
M82         & I0; HII         & 4.14 & 3.9 & 5       & 0112290201 & 29387 & Starburst \\
Holmberg IX & Im              & 4.0  & 3.6 & 4       & 0112521001 & 10350 & M81 also in field of view \\
Sextans A   & IBm             & 3.85 & 1.4 & 6       & 0026340201 & 21618 & - \\
IC 2574     & SAB(s)m         & 2.29 & 3.6 & 7       & 0026340301 & 24263 & bursting star-formation \\
NGC 4214    & IAB(s)m; HII    & 1.49 & 2.7 & \nodata & 0035940201 & 14744 & - \\
NGC 4258    & SAB(s)bc;LINER  & 1.2  & 7.2 & \nodata & 0059140901, 0110920101 & 16146, 21895 & - \\
NGC4395     & SA(s)m;LINER   & 1.33 & 4   & \nodata & 0112521901 & 15842 & - \\
NGC4449     & IBm; HII        & 1.39 & 3.08& \nodata & 0112521701 & 15522 & - \\
NGC4490     & SB(s)d          & 1.78 & 7.8 & 1       & 0112280201 & 17754 & interacting with NGC4485 \\
NGC4631     & SB(s)d          & 1.28 & 7.5 & 1       & 0110900201 & 53850 & - \\
NGC4736     &(R)SA(r)ab;LINER & 1.43 & 4.3 & 1       & 0094360601 & 23461 & - \\
NGC4945     & SB(s)cd; Sy2    & 15.9 & 3.1 & \nodata & 0112310301 & 23062 & - \\
NGC 5204    &  SA(s)m; HII    & 1.42 & 4.8 & 1      & 0142770101, 0142770301 & 19205, 16387 & - \\
M51         &  Sc; Sy2        & 1.55 & 7.2 & 1       & 0112840201 & 20924 & Galaxy pair \\
M83         & SAB(s)c;HII     & 3.94 & 6.2 & \nodata & 0110910201 & 30627 & Starburst \\
NGC5253     & Im pec;HII      & 3.77 & 3.2 & 1      & 0035940301 & 47216 & Starburst \\
M101        & SAB(rs)cd       & 1.17 & 7.4 & 8      & 0104260101 & 43019 & - \\
NGC5408     & IB(s)m; HII     & 5.73 & 4.8 & 9       & 0112290601 & 7757  & - \\
Circinus    & SA(s)b; Sy2     & 57.8 & 4   & 10      & 0111240101 & 110496& - \\
\enddata

\tablenotetext{a}{from the NASA/IPAC Extragalactic Database (NED)}
\tablenotetext{b}{column density in units of $10^{20}$\,cm$^{-2}$, obtained from the web version of the nH FTOOL}
\tablenotetext{c}{distance in Mpc (if no reference is given, obtained from the distance modulus given in LEDA)}
\tablenotetext{d}{{\it XMM-Newton} observation ids for the data examined in this survey}

\tablerefs{
(1)Tully 1988; (2) Tosi et al. 2001; (3) Tolstoy et al. 1995; (4) Freedman et al. 1994;
(5) Sakai \& Madore 1999; (6) Sakai, Madore, \& Freedman 1996; (7)Shapley, Fabbiano, \& Eskridge 2001;
(8) Kelson 1996; (9) Karachentsev et al. 2002; (10) Freeman et al. 1977.}

\end{deluxetable}

\clearpage

\begin{deluxetable}{lllllllll}
\tabletypesize{\scriptsize}
\tablecaption{{\it XMM-Newton} best fit: single component spectral fits\label{tbl-3}}
\tablewidth{0pt}
\tablehead{
\colhead{} & \multicolumn{3}{c}{Powerlaw}  & \multicolumn{3}{c}{Bremsstrahlung} &
\colhead{} & \colhead{} \\
\cline{2-4} \cline{5-7} \\
\colhead{Source} & \colhead{n$_H$\tablenotemark{a}} & \colhead{$\Gamma$} & \colhead{$\chi^{2}/$dof} &
\colhead{n$_H$\tablenotemark{a}} & \colhead{kT (keV)} & \colhead{$\chi^{2}/$dof} & \colhead{$F_X$\tablenotemark{b}} & 
\colhead{$L_X$\tablenotemark{c}}
}

\startdata
NGC247 XMM2   & $1.4^{+1.8} _{-1.1}$ & $2.29^{+1.02} _{-0.57}$ & {\bf 47.7/54} & $< 0.65$ & $2.55^{+6.90} _{-1.61}$ & 48.8/54 & 0.33 & 0.38 \\
NGC 253 XMM2 (obs 1) &$1.6^{+0.4} _{-0.3}$ & $2.51^{+0.18} _{-0.17}$ & {\bf 69.1/74} & $0.5^{+0.2} _{-0.3}$ & $2.12^{+0.52} _{-0.37}$ & 74.7/74 & 0.52 & 0.87 \\
NGC300 XMM4\tablenotemark{d}   & $2.5$ & $9.07$ & 90.6/45 & $0.27$ & $0.14$ & 117.6/45 & - & - \\
NGC1313 XMM4  & $1.86^{+0.5} _{-0.4}$& $1.8^{+0.07} _{-0.12}$  & {\bf 141.7/149} & $1.2^{+0.3} _{-0.3}$ & $6.62^{+2.3} _{-1.48}$ & 140.1/149 & 0.33 & 0.69 \\
IC0342 XMM1   & $5.8^{+0.6} _{-0.3}$ & $1.68^{+0.08} _{-0.08}$ & {\bf 159.5/185} & $4.9^{+0.5} _{-0.4}$ & $10.5^{+3.3} _{-1.9}$ & 160/185 & 3.5 & 6.37 \\
IC0342 XMM2   & $23.9^{+4.0} _{-3.6}$ & $1.85^{+0.22} _{-0.20}$ & {\bf 77.5/85} & $21^{+3.0} _{-2.8}$ & $8.5^{+5.0} _{-2.4}$ & 74.9/85 & 4.64 & 8.44 \\
IC0342 XMM4   & $5.3^{+1.4} _{-1.2}$ & $2.02^{+0.20} _{-0.19}$ & 64/58 & $4.2^{+0.99} _{-0.85}$ & $4.44^{+1.68} _{-0.74}$ & {\bf 56.9/58} & 0.69 & 1.26 \\
MRK71 XMM1    & $0.47^{+0.30} _{-0.32}$ & $1.69^{+0.11} _{-0.13}$ & {\bf 55.3/54} & $0.04^{+0.26} _{-0.04}$ & $7.98^{+4.90} _{-2.92}$ & 59/56 & 0.19 & 0.27 \\
NGC2403 XMM4 & $1.7^{+0.8} _{-0.7}$& $1.89^{+0.30} _{-0.25}$  & {\bf 62.3/71} & $1.1^{+0.5} _{-0.3}$ & $4.59^{+4.1} _{-1.5}$ & 62.3/71 & 0.31 & 0.48 \\
HolmII XMM1 (obs 2)  & $1.5^{+0.2} _{-0.2}$ & $3.09^{+0.15} _{-0.12}$ & {\bf 266.7/252} & $0.31^{+0.12} _{-0.15}$& $1.13^{+0.11} _{-0.11}$ & 309.4/252 & 3.5 & 3.1 \\
Holm I XMM2   & 0.35\tablenotemark{e} & $2.13^{+0.16} _{-0.15}$ & {\bf 39.2/45} & $0.35\tablenotemark{e}$& $2.0^{+0.57} _{-0.48}$ & 68.8/45 & 0.10 & 0.16 \\
Holm I XMM3   & 0.35\tablenotemark{e} & $2.05^{+0.19} _{-0.18}$ & {\bf 34.4/32} &0.35\tablenotemark{e} & $2.03^{+0.85} _{-0.56}$ & 42.1/32 & 0.12 & 0.19 \\
IC2574 XMM1   & $1.3^{+0.40} _{-0.30}$ & $1.97^{+0.07} _{-0.10}$ & 120.9/103 & $0.69^{+0.23} _{-0.25}$& $4.1^{+0.89} _{-0.67}$ & {\bf 107.5/103} & 0.35 & 0.47 \\
IC2574 XMM2   & $0.4^{+0.4} _{-0.3}$ & $2.2^{+0.21} _{-0.09}$ & {\bf 45.7/51} & $0.229\tablenotemark{e}$& $1.71^{+0.33} _{-0.27}$ & 65.4/52 & 0.22 & 0.34 \\
IC2574 XMM3   &$0.15^{+0.35} _{-0.14}$& $2.43^{+0.27} _{-0.18}$ & {\bf 40.3/49} & 0.229\tablenotemark{e}& $0.97^{+0.18} _{-0.14}$ & 76.5/49 & 0.22 & 0.34 \\
NGC4214 XMM1  & $1.1^{+0.52} _{-0.47}$ & $1.87^{+0.26} _{-0.21}$ & {\bf 41.9/38} & $0.54^{+0.41} _{-0.35}$& $4.86^{+4.52} _{-1.66}$ & 44.5/38 & 0.25 & 0.22 \\
NGC4258 XMM2 (obs 2)& $6.7^{+2.6} _{-1.5}$ & $2.49^{+0.44} _{-0.33}$ & {\bf 83.6/57} & $4.8^{+0.9} _{-1.3}$& $2.61^{+1.22} _{-0.72}$ & 85.5/57 & 0.30 & 1.9 \\ 
NGC4258 XMM3  & $1.4^{+0.69} _{-0.64}$ & $2.32^{+0.34} _{-0.24}$ & {\bf 38.9/37} & $0.49^{+0.44} _{-0.38}$& $2.48^{+1.09} _{-0.74}$ & 41.3/37 & 0.20 & 1.2 \\
\nodata       & 3.8 & 1.82 & {\bf 4/11} & 2.7& 7.14 & 5/11 & 0.077 & 0.48 \\ 
NGC4258 XMM4  & $0.68^{+0.24} _{-0.42}$ & $1.97^{+0.22} _{-0.19}$ & {\bf 41.1/48} & $0.06^{+0.31} _{-0.05}$& $4.07^{+1.6} _{-1.2}$ & 45.2/48 & 0.39 & 2.4 \\
\nodata       & $1.9^{+0.78} _{-0.60}$ & $2.24^{+0.29} _{-0.24}$ & {\bf 77.03/77} & $0.9^{+0.6} _{-0.4}$& $2.82^{+1.2} _{-0.8}$ & 77.8/77 & 0.33 & 2.0 \\ 
NGC4395 XMM2  & $0.33^{+0.6} _{-0.3}$  & $2.75^{+0.45} _{-0.33}$ & {\bf 38.6/36} & $0.133\tablenotemark{e}$& $0.79^{+0.13} _{-0.13}$ & 52/37 & 0.15 & 0.28 \\
NGC4395 XMM4  & $0.3^{+0.6} _{-0.3}$  & $2.08^{+0.39} _{-0.30}$ & {\bf 16/25} & $0.133\tablenotemark{e}$& $1.97^{+0.99} _{-0.60}$ & 21.6/26 & 0.15 & 0.28 \\
NGC4449 XMM2  & $1.5^{+0.3} _{-0.3}$  & $2.81^{+0.16} _{-0.14}$ & {\bf 103.5/112} & $0.25^{+0.2} _{-0.2}$& $1.65^{+0.22} _{-0.21}$ & 112.1/112 & 0.29 & 0.33 \\
NGC4490 XMM4  & $10.2^{+2.3} _{-1.8}$  & $2.09^{+0.23} _{-0.19}$ & {\it 51.6/50} & $8.3^{+1.3} _{-1.5}$& $4.75^{+1.82} _{-0.90}$ & 50.3/50 & 0.84 & 6.1 \\
NGC4490 XMM5  & $3.9^{+0.94} _{-0.81}$  & $2.31^{+0.22} _{-0.20}$ & {\bf 60.1/65} & $2.5^{+0.54} _{-0.59}$& $3.08^{+0.89} _{-0.62}$ & 61.6/65 & 0.41 & 2.98 \\
NGC4631 XMM4\tablenotemark{d}  & $7.8$  & $9.50$ & 261.5/74 & $2.9$& $0.17$ & 207.8/74 & - & - \\
NGC4631 XMM5\tablenotemark{f}  & $1.3$  & $1.03$ & 641.8/153 & $1.3$& $199$ & 659/153 & - & - \\
NGC4945 XMM3  & $3.3^{+1.3} _{-0.9}$ & $1.82^{+0.12} _{-0.20}$ & {\bf 30.1/30} & $2.5^{+0.83} _{-0.90}$ & $6.07^{+4.50} _{-1.71}$ & 30.3/30 & 0.38 & 0.43 \\
NGC5204 XMM2  & $0.89^{+0.49} _{-0.53}$ & $1.98^{+0.25} _{-0.20}$ & {\it 42.37/42} & $0.23^{+0.3} _{-0.22}$& $4.05^{+1.51} _{-0.95}$ & 42.2/42 & 0.15 & 0.41 \\
\nodata       & $0.75^{+0.45} _{-0.45}$ & $1.63^{+0.20} _{-0.17}$ & {\bf 41.4/47} & $0.42^{+0.46} _{-0.38}$& $7.82^{+5.03} _{-2.70}$ & 39.4/47 & 0.25 & 0.69 \\
M51 XMM3      & $0.6^{+0.30} _{-0.40}$ & $1.86^{+0.09} _{-0.15}$ & {\bf 63.2/72} & $0.05^{+0.3} _{-0.02}$& $5.22^{+2.26} _{-1.41}$ & 69.2/72 & 0.18 & 1.1 \\
M51 XMM4      & $0.4^{+0.20} _{-0.30}$ & $1.55^{+0.08} _{-0.13}$ & {\bf 34.8/37} & $0.01^{+0.17} _{-0.13}$& $11.1^{+0.32} _{-0.25}$ & 34.8/37 & 0.16 & 0.99 \\
\enddata

\tablenotetext{a}{total column density in units of $10^{21}$\,cm$^{-2}$}
\tablenotetext{b}{unabsorbed flux in the 0.3-10 keV band in units of $10^{-12}$\,erg\,cm$^{-2}$\,s$^{-1}$}
\tablenotetext{c}{unabsorbed luminosity in the 0.3-10 keV band, using the distances quoted in Table~\ref{tbl-1}, in units of $10^{39}$\,erg\,s$^{-1}$}
\tablenotetext{d}{see appendix; super-soft X-ray source best fit by single-component blackbody}
\tablenotetext{e}{absorbtion column density fixed to the galactic column density found in Table~\ref{tbl-1}}
\tablenotetext{f}{source is best fit by a combined power law and {\it vapec} model; see appendix}
\end{deluxetable}

\clearpage

\begin{deluxetable}{llllllll}
\tabletypesize{\scriptsize}

\tablecaption{{\it XMM-Newton} best fit: two-component blackbody and power law spectral fits\label{tbl-4}}
\tablewidth{0pt}
\tablehead{
\colhead{Source} & \colhead{n$_H$\tablenotemark{a}} & \colhead{kT (keV)} & \colhead{$\Gamma$} & \colhead{$\chi^{2}/$dof} &
\colhead{$\Delta\chi^{2}$\tablenotemark{b}} &\colhead{$F_X$\tablenotemark{c}} & \colhead{$L_X$\tablenotemark{d}}
}

\startdata
NGC247 XMM1   & $4.1^{+1.9} _{-1.5}$ & $0.12^{+0.03} _{-0.02}$ &$4.18^{+1.79} _{-2.52}$ & 86.5/93 & 25.7 & 6.2 & 7.1 \\
NGC253 XMM1   & $2.7^{+0.4} _{-0.4}$ & $0.80^{+0.12} _{-0.09}$ &$1.74^{+0.17} _{-0.14}$ & 225.9/230 & 36.7 & 2.7 & 4.5 \\
\nodata       & $7.3^{+1.1} _{-0.9}$ & $1.14^{+0.07} _{-0.10}$ &$2.54^{+0.27} _{-0.22}$ & 567/580 & 44.6 & 3.4 & 5.7 \\ 
NGC253 XMM2 (obs 2)  & $2.0^{+0.3} _{-0.2}$ & $0.71^{+0.10} _{-0.10}$ &$2.14^{+0.05} _{-0.08}$ & 460.3/498 & 47.1 & 1.6 & 2.7 \\
NGC253 XMM3   & $3.1^{+4.8} _{-0.5}$ & $0.75^{+0.13} _{-0.10}$ &$2.47^{+2.99} _{-0.41}$ & 68.5/82 & 23.4 & 0.60 & 1.0 \\
\nodata       & $3.2^{+0.7} _{-0.5}$ & $0.67^{+0.13} _{-0.09}$ &$2.07^{+0.14} _{-0.20}$ & 347.4/407 & 34.4 & 0.80 & 1.3 \\
NGC253 XMM4   & $20^{+10.8} _{-7.6}$ & $0.11^{+0.03} _{-0.03}$ &$2.51^{+0.49} _{-0.30}$ & 66.7/57 & 6.9 & 15 & 25 \\
\nodata       & $4.5^{+1.2} _{-1.9}$ & $0.09^{+0.02} _{-0.01}$ &$2.33^{+0.27} _{-0.22}$ & 309.3/291 & 12.1 & 1.4 & 2.2 \\
NGC253 XMM5   & $1.5^{+7.2} _{-1.5}$ & $0.96^{+0.24} _{-0.32}$ &$2.43^{+3.06} _{-1.36}$ & 26.5/23 & 5.3 & 0.26 & 0.43 \\
\nodata       & $4.6^{+1.1} _{-0.7}$ & $0.16^{+0.02} _{-0.03}$ &$1.95^{+0.14} _{-0.11}$ & 223.7/296 & 60.1 & 1.4 & 2.2 \\
NGC253 XMM6   & $6.3^{+2.1} _{-1.1}$ & $0.12^{+0.02} _{-0.02}$ &$2.26^{+0.18} _{-0.12}$ & 417.9/407 & 17.1 & 1.9 & 3.1 \\
NGC253 XMM7   & $6.3^{+0.9} _{-1.1}$ & $0.69^{+0.11} _{-0.12}$ &$2.40^{+0.41} _{-0.17}$ & 335.8/339 & 21.2 & 1.1 & 1.8 \\
NGC300 XMM1   & $1.7^{+0.20} _{-0.30}$ & $0.98^{+0.14} _{-0.10}$ &$3.41^{+0.06} _{-0.26}$ & 443.7/420 & 26.1 & 1.3 & 1.0 \\
NGC300 XMM2   & $3.8^{+1.7} _{-1.4}$ & $0.09^{+0.01} _{-0.01}$ &$2.87^{+0.34} _{-0.38}$ & 102.6/97 & 31.34 & 1.1 & 0.86 \\
NGC300 XMM3   & $4.4^{+1.0} _{-0.8}$ & $0.04^{+0.25} _{-0.01}$ &$1.98^{+0.1} _{-0.1}$ & 87.7/79 & 14.2 & 1.2 & 0.93 \\
NGC300 XMM6   & $2.3^{+2.6} _{-1.3}$ & $0.84^{+0.25} _{-0.19}$ &$4.9^{+1.97} _{-0.7}$ & 34.6/35 & 13 & 0.27 & 0.20 \\
NGC1313 XMM1  & $3.0^{+1.2} _{-0.9}$ & $0.13^{+0.03} _{-0.02}$ &$1.75^{+0.14} _{-0.11}$ & 194.1/201 & 35.4 & 0.64 & 1.3 \\
NGC1313 XMM2  & $3.1^{+0.4} _{-0.3}$ & $0.16^{+0.04} _{-0.02}$ &$2.27^{+0.10} _{-0.14}$ & 425.2/419 & 38.9 & 2.0 & 4.2 \\
NGC1313 XMM3  & $6.2^{+0.8} _{-0.6}$ & $0.11^{+0.01} _{-0.01}$ &$2.76^{+0.10} _{-0.11}$ & 441.7/424 & 336.6 & 10 & 22 \\
IC0342 XMM3   & $9.7^{+1.8} _{-2.1}$ & $0.09^{+0.02} _{-0.01}$ &$2.69^{+0.16} _{-0.23}$ & 129.5/107 & 56.3 & 31 & 56.4 \\
NGC1705 XMM1  & $0.29^{+0.39} _{-0.27}$ & $1.01^{+0.41} _{-0.29}$ &$2.31^{+0.89} _{-0.48}$ & 53/85 & 8.9 & 0.10 & 0.44 \\
NGC1705 XMM3  & $< 1.44$ & $1.07^{+0.20} _{-0.15}$ &$2.23^{+0.70} _{-0.56}$ & 69.8/65 & 11.1 & 0.15 & 0.48 \\
NGC2403 XMM1  & $2.3^{+1.2} _{-1.1}$ & $0.66^{+0.16} _{-0.18}$ &$2.18^{+0.41} _{-0.59}$ & 81.4/79 & 10.8 & 1.99 & 3.1 \\
NGC2403 XMM2  & $1.8^{+0.8} _{-0.6}$ & $0.62^{+0.16} _{-0.11}$ &$1.95^{+0.26} _{-0.42}$ & 163.1/151 & 16.4 & 1.0 & 1.6 \\
NGC2403 XMM3  & $1.7^{+1.1} _{-0.8}$ & $0.74^{+0.23} _{-0.21}$ &$2.15^{+0.66} _{-0.40}$ & 84.2/105 & 8.4 & 0.64 & 1.1 \\
Holm II XMM1(obs 1) & $1.6^{+0.1} _{-0.2}$ & $0.14^{+0.02} _{-0.01}$ &$2.35^{+0.05} _{-0.11}$ & 997.5/976 & 136.7 & 12 & 10 \\
M81 XMM1      & $3.3^{+0.17} _{-0.08}$ & $0.90^{+0.03} _{-0.03}$ &$2.52^{+0.03} _{-0.04}$ & 1273.7/1243 & 533.1 & 4.5 & 7.0 \\
\nodata       & $3.5^{+0.4} _{-0.6}$ & $1.13^{+0.13} _{-0.14}$ &$2.34^{+0.29} _{-0.36}$ & 203.5/204 & 21.4 & 4.8 & 7.4 \\
M81 XMM2      & $7.4^{+0.5} _{-0.7}$ & $0.1^{+0.004} _{-0.004}$ &$2.87^{+0.16} _{-0.17}$ & 833.9/616 & 524.3 & 13 & 22 \\ 
M81 XMM4      & $1.1^{+1.6} _{-1.0}$ & $2.51^{+1.11} _{-0.73}$ &$2.31^{+1.22} _{-1.05}$ & 48.9/50 & 28.2 & 0.43 & 0.70 \\ 
M81 XMM5      & $0.15^{+0.69} _{-0.13}$ & $0.62^{+0.19} _{-0.11}$ &$1.26^{+0.22} _{-0.20}$ & 89/80 & 8.5 & 0.38 & 0.59 \\ 
Holm IX XMM1  & $2.1^{+0.2} _{-0.2}$ & $0.17^{+0.02} _{-0.02}$ &$1.72^{+0.04} _{-0.03}$ & 866.6/878 & 134.3 & 10 & 16 \\
NGC4258 XMM1  & $0.38^{+0.96} _{-0.3}$ & $0.54^{+0.17} _{-0.08}$ &$1.51^{+0.4} _{-0.4}$ & 91.1/76 & 10.3 & 0.34 & 2.1 \\
NGC4258 XMM2 (obs 1) & $1.9^{+2.4} _{-0.4}$ & $0.78^{+0.12} _{-0.13}$ &$2.02^{+0.65} _{-1.8}$ & 73.4/61 & 24.1 & 0.31 & 1.9 \\
NGC4395 XMM1  & $2.0^{+0.08} _{-0.07}$ & $0.14^{+0.02} _{-0.02}$ &$3.44^{+0.54} _{-0.56}$ & 168.2/154 & 26.9 & 1.4 & 2.7 \\
NGC4449 XMM1  & $8.7^{+4.8} _{-2.1}$ & $0.19^{+0.13} _{-0.07}$ &$2.21^{+0.33} _{-0.29}$ & 111.2/116 & 4.3 & 1.56 & 1.8 \\ 
NGC4449 XMM3  & $3.5^{+1.3} _{-0.9}$ & $0.15^{+0.03} _{-0.03}$ &$2.52^{+0.36} _{-0.39}$ & 119.9/87 & 34.1 & 1.1 & 1.3 \\ 
NGC4490 XMM1  & $5.8^{+2.96} _{-2.96}$ & $0.77^{+0.08} _{-0.095}$ &$2.89^{+1.77} _{-0.61}$ & 66.5/63 & 35 & 0.88 & 6.4 \\ 
NGC4631 XMM1  & $3^{+0.9} _{-0.5}$ & $0.12^{+0.03} _{-0.02}$ &$2.12^{+0.03} _{-0.02}$ & 371.3/345 & 12.1 & 0.96 & 6.5 \\ 
NGC4631 XMM2  & $2.3^{+1.4} _{-0.3}$ & $0.18^{+0.05} _{-0.06}$ &$1.80^{+0.12} _{-0.09}$ & 107.4/97 & 12.1 & 0.25 & 1.7 \\
NGC4631 XMM3  & $1.1^{+1.1} _{-0.8}$ & $1.01^{+0.12} _{-0.1}$ &$2.45^{+1} _{-0.62}$ & 127.1/96 & 18.9 & 0.15 & 1.0 \\
NGC4945 XMM1  & $3.5^{+2.1} _{-1.1}$ & $0.77^{+0.27} _{-0.10}$ &$1.60^{+0.40} _{-0.31}$ & 96.1/120 & 20 & 0.59 & 0.68 \\  
NGC4945 XMM2  & $3.2^{+1.1} _{-0.7}$ & $1.15^{+0.28} _{-0.33}$ &$1.80^{+0.20} _{-0.30}$ & 105.8/113 & 8.7 & 0.66 & 0.76 \\  
NGC4945 XMM4  & $4.0^{+2.0} _{-1.1}$ & $0.61^{+0.10} _{-0.10}$ &$2.82^{+1.06} _{-0.58}$ & 58.4/60 & 17.1 & 0.38 & 0.44 \\
NGC5204 XMM1  & $0.66^{+0.35} _{-0.08}$ & $0.16^{+0.02} _{-0.03}$ &$1.92^{+0.12} _{-0.06}$ & 543.0/559 & 49.1 & 1.98 & 5.6 \\
\nodata       & $1.1^{+0.08} _{-0.14}$ & $0.16^{+0.02} _{-0.02}$  &$2.03^{+0.12} _{-0.12}$ & 461.4/496 & 71.6 & 2.92 & 8.0 \\
M51 XMM1      & $0.95^{+1.10} _{-0.18}$ & $0.16^{+0.03} _{-0.05}$  &$2.15^{+0.42} _{-0.17}$ & 97/80 & 13.4 & 0.31 & 1.9 \\
M51 XMM5      & $10.4^{+1.7} _{-3.7}$ & $0.078^{+0.01} _{-0.01}$ &$2.26^{+0.26} _{-0.25}$ & 59.8/70 & 196.2 & 220 & 1900 \\
M83 XMM4      & $1.77^{+3.9} _{-1.77}$ & $0.54^{+0.18} _{-0.09}$ &$1.61^{+0.96} _{-0.31}$ & 84.8/89 & 6.6 & 0.2 & 0.92 \\
M101 XMM1     & $0.22^{+0.12} _{-0.15}$ & $0.21^{+0.03} _{-0.04}$ &$1.42^{+0.14} _{-0.05}$ & 249.9/231 & 53.1 & 0.45 & 2.9 \\
M101 XMM2     & $1.6^{+0.46} _{-0.21}$ & $0.76^{+0.14} _{-0.10}$ &$1.88^{+0.25} _{-0.11}$ & 251.6/261 & 37.2 & 0.7 & 4.6 \\
NGC5408 XMM1  & $0.9^{+0.21} _{-0.16}$ & $0.14^{+0.01} _{-0.01}$ &$2.71^{+0.16} _{-0.20}$ & 316.4/337 & 80.4 & 3.97 & 10.9 \\
CIRCINUS XMM1 & $10.1^{+1.2} _{-1.2}$ & $0.10^{+0.01} _{-0.01}$ &$2.30^{+0.08} _{-0.08}$ & 749.4/861 & 13.5 & 12 & 23 \\
CIRCINUS XMM2 & $11.2^{+2.4} _{-1.7}$ & $0.53^{+0.03} _{-0.03}$ &$4.71^{+0.94} _{-0.49}$ & 438.5/430 & 79.4 & 5.6 & 10.7 \\
CIRCINUS XMM3 & $13.5^{+5.5} _{-5.6}$ & $0.67^{+0.10} _{-0.08}$ &$5.77^{+2.24} _{-2.3}$ & 269.3/260 & 15.9 & 7.6 & 14.5 \\
\enddata

\tablenotetext{a}{total column density in units of $10^{21}$ cm$^{-2}$}
\tablenotetext{b}{improvement in $\chi^2$ over the single-component power law model}
\tablenotetext{c}{unabsorbed flux in the 0.3-10 keV band in units of $10^{-12}$ erg cm$^{-2}$ s$^{-1}$}
\tablenotetext{d}{unabsorbed luminosity in the 0.3-10 keV band, using the distances quoted in Table~\ref{tbl-1}, in units of $10^{39}$ erg s$^{-1}$}
\end{deluxetable}

\clearpage

\begin{deluxetable}{llllllll}
\tabletypesize{\scriptsize}

\tablecaption{{\it XMM-Newton} two-component blackbody and power law spectral fits for sources with large uncertainty\label{tbl-13}}
\tablewidth{0pt}
\tablehead{
\colhead{Source} & \colhead{n$_H$\tablenotemark{a}} & \colhead{kT (keV)} & \colhead{$\Gamma$} & \colhead{$\chi^{2}/$dof} &
\colhead{$\Delta\chi^{2}$\tablenotemark{b}} &\colhead{$F_X$\tablenotemark{c}} & \colhead{$L_X$\tablenotemark{d}}
}

\startdata
NGC300 XMM5   &$0.41^{+0.60} _{-0.30}$& $1.06^{+0.37} _{-0.20}$&$2.78^{+0.61} _{-0.65}$& 46.6/53& 7.6    &0.17 & 0.13 \\
NGC1705 XMM2  & $0.96^{+0.97} _{-0.32}$ & $0.23^{+0.10} _{-0.11}$ &$1.60^{+1.97} _{-0.27}$ & 85.5/74 & 6.5 & 0.09 & 0.27 \\
Holm I XMM1   & $0.4^{+0.5} _{-0.3}$ & $1.97^{+0.66} _{-0.89}$ &$2.46^{+0.44} _{-0.40}$ & 97.4/93 & 5.4 & 0.6 & 0.93 \\
M81 XMM3      & $3.7^{+2.4} _{-2.1}$ & $0.11^{+0.05} _{-0.02}$ &$1.69^{+0.27} _{-0.33}$ & 77.1/78 & 4.25 & 0.53 & 0.82 \\ 
Sextans A XMM1& $0.4^{+0.7} _{-0.1}$ & $1.05^{+2.3} _{-0.07}$ & $2.6^{+0.8} _{-0.2}$ & 269.1/271    & 2.3 &0.60 & 0.14 \\
NGC4214 XMM2  & $1.8^{+1.9} _{-0.6}$ & $0.81^{+0.56} _{-0.21}$ &$3.95^{+1.81} _{-1.05}$ & 46.4/44 & 4.5 & 0.4 & 0.35 \\  
NGC4395 XMM3  & $0.5^{+0.9} _{-0.3}$ & $1.10^{+0.67} _{-0.18}$ &$2.66^{+1.05} _{-0.77}$ & 52/56 & 3.9 & 0.29 & 0.56 \\
NGC4490 XMM2  & $4.4^{+1.9} _{-1.9}$ & $0.60^{+0.20} _{-0.12}$ &$2.13^{+0.50} _{-0.70}$ & 42.4/54 & 7.1 & 0.65 & 4.7 \\ 
NGC4490 XMM3  & $13^{+9.6} _{-2.5}$ & $0.09^{+0.02} _{-0.02}$ &$3.21^{+0.52} _{-0.17}$ & 72.1/78 & 4.6 & 12 & 87.4 \\  
NGC4736 XMM1  & $6.3^{+3.0} _{-3.7}$ & $0.08^{+0.03} _{-0.02}$ & $2.41^{+0.34} _{-0.27}$ & 54.9/51 & 7.9& 8.1 & 17.9 \\
M51 XMM2      & $1.3^{+0.6} _{-0.5}$ & $0.26^{+0.07} _{-0.08}$ &$1.80^{+0.61} _{-0.92}$ & 70.7/68 & 4.5 & 0.36 & 3.0 \\
M51 XMM6      &$8.2^{+3.5} _{-5.6}$& $0.08^{+0.05} _{-0.02}$ & $3.0^{+0.37} _{-0.43}$ &36.9/41 & 4.07& 5.6 & 35 \\
M51 XMM7      & $2.8^{+3.4} _{-2.1}$ & $0.10^{+0.03} _{-0.03}$ &$1.97^{+0.43} _{-0.30}$ & 31.7/29 & 6.1 & 0.26 & 1.6 \\
M83 XMM1      & $1.6^{+0.48} _{-0.45}$ & $0.74^{+0.23} _{-0.26}$ &$2.58^{+0.60} _{-0.24}$ & 177.7/209 & 4.7 & 0.63 & 2.5 \\
M101 XMM3     &$1.98^{+1.0} _{-0.61}$& $0.63^{+0.22} _{-0.20}$ & $2.93^{+0.15} _{-0.26}$ &145.5/131 & 3.4& 0.56 & 3.7 \\
M101 XMM4     & $1.8^{+0.17} _{-0.15}$ & $0.54^{+0.11} _{-0.07}$ &$2.22^{+0.12} _{-0.08}$ & 158.2/138 & 7.5 & 0.34 & 2.2 \\
M101 XMM5     &$1.3^{+1.2} _{-0.2}$& $0.18^{+0.05} _{-0.06}$ & $1.95^{+0.3} _{-0.22}$ &45.1/44 & 2.8 & 0.13 & 0.85 \\
\enddata

\tablenotetext{a}{total column density in units of $10^{21}$ cm$^{-2}$}
\tablenotetext{b}{improvement in $\chi^2$ over the single-component power law model}
\tablenotetext{c}{unabsorbed flux in the 0.3-10 keV band in units of $10^{-12}$ erg cm$^{-2}$ s$^{-1}$}
\tablenotetext{d}{unabsorbed luminosity in the 0.3-10 keV band, using the distances quoted in Table~\ref{tbl-1}, in units of $10^{39}$ erg s$^{-1}$}
\end{deluxetable}

\clearpage

\begin{deluxetable}{lllll}
\tabletypesize{\scriptsize}
\tablecaption{Bolometric Luminosities of ULX sources\label{tbl-11}}
\tablewidth{0pt}
\tablehead{
\colhead{Source} & \colhead{L$_{cutoffpl}$\tablenotemark{a}} & \colhead{L$_{bol}$}\tablenotemark{b} & 
\colhead{M$_{Edd}$\tablenotemark{c}} & \colhead{class\tablenotemark{d}}
}

\startdata
NGC247 XMM1	& 13.4258	& 7.07734	& 54 	& HS ULX\\
NGC253 XMM1	& 9.31469	& 2.44574	& 19 	& HS ULX\\
NGC253 XMM2	& 4.3701	& 2.15292	& 17 	& HS ULX\\
NGC253 XMM6	& 5.05828	& 3.92514	& 30 	& HS ULX\\
NGC1313 XMM3	& 37.0364	& 27.9692	& 215 	& HS ULX\\
NGC1313 XMM4	& 1.50345	& \nodata	& 116 	& LS IMBH cand.\\
IC0342 XMM1	& 14.1215	& \nodata	& 1086 	& LS ULX\\
IC0342 XMM2	& 19.8129	& \nodata	& 1524 	& LS ULX\\
IC0342 XMM3	& 114.015	& 95.4068	& 734 	& HS ULX\\
MRK71 XMM1      & 0.2993	& \nodata	& 23 	& LS IMBH cand.\\
NGC2403 XMM1	& 4.1497	& 2.14873	& 17 	& HS ULX\\
NGC2403 XMM4	& 0.57068	& \nodata	& 44 	& LS IMBH cand.\\
Holmberg II XMM1& 0.88906	& \nodata	& 68 	& LS IMBH cand.\\
\nodata		& 16.8335	& 11.4543	&  88 	& HS ULX\\
Holmberg I XMM2 & 10.5158	& \nodata	& 809 	& LS ULX\\
M81 XMM1	& 15.7004	& 3.17932	& 24 	& HS ULX\\
Holmberg IX XMM1& 31.0582	& 28.1445	& 216 	& HS ULX\\
NGC4214 XMM1	& 0.26699	& \nodata	& 21 	& LS IMBH cand.\\
NGC4258 XMM3	& 0.46503	& \nodata	& 36 	& LS IMBH cand.\\
NGC4395 XMM1	&  9.04609	& 2.94683	& 23 	& HS ULX\\
NGC4449 XMM2	& 2.48586	& \nodata	& 191 	& LS IMBH cand.\\
NGC4490 XMM1	& 16.8513	& 3.21972	& 25  	& HS ULX\\
NGC4490 XMM2	& 7.36612	& 4.51554	&  35 	& HS ULX\\
NGC4490 XMM3	& 240.653	& 176.04	& 1354 	& HS ULX\\
NGC4490 XMM4	& 1.66829	& \nodata	& 128 	& LS IMBH cand.\\
NGC4490 XMM5	& 12.7136	& \nodata	& 978 	& LS ULX\\
NGC4631 XMM1	& 10.6527	& 8.59661	& 66 	& HS ULX\\
NGC4736 XMM1	& 31.6561	& 27.3664	& 211 	& HS ULX\\
NGC4945 XMM3	& 0.44985	& \nodata	& 35	& LS IMBH cand. \\
NGC5204 XMM1	& 22.4756	& 2.20492	& 17 	& HS ULX\\
NGC5204 XMM2	& 5.57769	& \nodata	& 429 	& LS ULX\\
M51 XMM2	& 4.25898	& 3.57502	& 28 	& HS ULX\\
M51 XMM3	& 2.10133	& \nodata	& 162 	& LS IMBH cand.\\
M51 XMM4	& 2.56064	& \nodata	& 197 	& LS IMBH cand.\\
M51 XMM6	& 46.7642	& 39.5189	& 304 	& HS ULX\\
M101 XMM1	& 8.04916	& 7.68224	& 59 	& HS ULX\\
M101 XMM2	& 7.54268	& 4.96709	& 38 	& HS ULX\\
M101 XMM3	& 10.9792	& 1.03659	& 8 	& HS ULX\\
NGC5408 XMM1	& 20.9211	& 11.5369	& 89 	& HS ULX\\
Circinus XMM1	& 70.3579	& 56.1033	& 432 	& HS ULX\\
Circinus XMM2	& 208.746	& 0.69929	& 5 	& HS ULX\\
Circinus XMM3	& 771.157	& 0.212957	& 2 	& HS ULX\\
\enddata
\tablenotetext{a}{estimation of the bolometric luminosity, determined with an exponential cut-off in the power law at high energy (see text)}
\tablenotetext{b}{bolometric luminosity estimate for high-state ULXs where the power law is cut at twice kT$_{in}$ (see text); units for both
luminosity measurements in $10^{39}$ erg s$^{-1}$}
\tablenotetext{c}{mass computed for objects radiating at $0.1 \times L_{Edd}$ (low-state objects) or $L_{Edd}$ (high-state objects; using L$_{bol}$), 
in units of M$_{\sun}$}
\tablenotetext{d}{classification based on the criteria set forward in the text: high-state ULX (HS ULX), low-state
ULX (LS ULX), and low-state IMBH candidate (low-state object with $ 10^{38}$\,erg\,s$^{-1} <$ L$_{bol} < 3 \times 10^{39}$\,erg\,s$^{-1}$)}
\end{deluxetable}

\clearpage

\clearpage 

\clearpage 

\clearpage 

\clearpage 

\clearpage 

\clearpage

\appendix
\section{Spectral Simulations}
\subsection{Two-component Model}
In order to determine the number of counts required to distinguish
whether a blackbody component is statistically significant for the
sources fit with a two-component model, we simulated spectra based on
that of some of the brightest sources.  We chose to simulate spectra
of bright two-component spectra exhibiting three different cases: (1)
the flux from the blackbody dominates over the power law component at
2\,kT, (2) an intermediary case, and (3) the flux from the power law
dominates over the blackbody component at 2\,kT .  Such simulations
would allow us to determine the uncertainty in our claims of a
combined fit being a better descriptor of the data.  This is necessary
because there is no a priori model which predicts the relative
intensities of the two components and, as we know from studies of
Galactic black holes, these components show a wide variety of relative
intensities.  To this end, we simulated spectra using the best-fit
absorbed blackbody and power law model with the {\it fakeit} command
in XSPEC.  We chose (1) NGC 247 XMM1, (2) NGC 5408 XMM1, and (3)
Holmberg II XMM1 as our seed observations.  These objects all have
very high signal-to-noise and thus the fits are robust.  The
respective ratios of powerlaw flux to blackbody flux contributions at
2\,kT are: (1) $<< 1.0$, (2) 1.77, and (3) 3.52.  All of these sources
have comparable blackbody temperatures indicative of our high-state
ULX candidates (roughly kT $\approx 0.15$).

We simulated 500 spectra each, using the two-component model, for each
of 2000\,counts, 1000\,counts, 400\,counts, and 200\,counts for the PN.  Each
simulated spectrum, based on the best-fit blackbody and power law
model, was fit with an absorbed blackbody and powerlaw model as well
as an absorbed pure-powerlaw model. We placed the constraint that the
blackbody temperature must lie within the range of $0.07 - 4.0$\,keV
(the range at which it would be detectable in the {\it XMM-Newton}
bandpass).  We allowed the power law index to vary over the range $0 -
4$ for the power law component of the combined blackbody and power law
model.  However, we placed a constraint that the power law component
must lie within the range $\Gamma = 1.5 - 2.0$ for the simple power
law model to be consistent with our fits to the sources we claim are
best fit by simple power laws.  This constraint ensures that the
spectral index would exhibit that of our classified ``low-state''
objects.

When analysing and classifying our real spectra, we declared a
detection of the thermal disk component if the addition of this
component (to a baseline powerlaw model) led to an improvement of the
goodness of fit parameter by at least $\Delta\chi^2=8$.  We can use
the above simulations to address the detectability of a thermal disk
component using this $\Delta\chi^2$ threshold as a function of the
relative strength of the thermal component and the number of counts in
the spectrum.  For each simulation, we fit the spectrum with both a
single absorbed power-law and a 2-component powerlaw and thermal disk
model and compute the quantity $\Delta\chi^2=\chi^2_{\rm
pow}-\chi^2_{\rm pow+disk}$.  In Fig.~A1, we plot the distribution of
$\Delta\chi^2$ from our 500 simulations for the weak  and strong
 blackbody component for spectra with 400\,counts and 2000\,counts.
It is clear that we cannot detect a weak thermal component in a 400
count spectrum --- the vast majority of the simulations ($\approx 82$\%) result in
$\Delta\chi^2<8$.  However, even a weak thermal component is easily
detected in a 2000 count spectrum (not a single simulation gave
$\Delta\chi^2<8$).  The strong blackbody case is
detectable with high significance even in a 400 count spectrum ($< 1$\%
of the simulations resulted in $\Delta\chi^2>8$).

When we increase the upper limit of the range of the spectral index in
{\it XSPEC} for the power law model to $\Gamma = 3.0$, our confidence
levels decrease.  For a weak thermal component with 400\,counts, all of the
simulations result in $\Delta\chi^2<8$.  At 2000\,counts, only $\approx 30$\%
of the simulations for a weak thermal component have $\Delta\chi^2<8$.
For a strong thermal component, 20\% of the simulations yield $\Delta\chi^2<8$ for
400\,counts while, as was the case for $\Gamma = 2.0$, none of the simulations
gave $\Delta\chi^2<8$ for 2000\,counts.   

Thus allowing
 the $\Gamma$ to ``float freely'' or remain unconstrained further decreases 
the confidence levels.  When the $\Gamma$  parameter is allowed to float the
 spectra are fit with higher $\Gamma$ values in order to compensate for
 the missing blackbody component.  When the upper limit was instituted
 at 2.0 or 3.0, we found that all 500 simulated spectra were fit with a
 $\Gamma = 2.0$ or 3.0, respectively, for the simple power law model.
 This same affect is not seen in the $\Gamma$ of the two-component
 model, where the value ranges between 1.4 and 4.0 with a peak in the
 distribution near that of the original model used to simulate the
 spectra.  Thus, higher power law indices ($> 3.0$) can indicate the
 necessity of an added blackbody component.  Since there are no
 Galactic black holes whose broad band spectra are well fit by steep
 power laws it seems that restricting the allowed power law indices
is more consistent with the nature of Galactic black holes.  In fitting our sources with the three ``standard''
 models, we allowed the $\Gamma$ to float, thus the problem of a
 missing blackbody component being compensated for by a steep power law
 should not have factored into our classification criteria.

\renewcommand{\thefigure}{A\arabic{figure}}


\begin{figure*}
\plotone{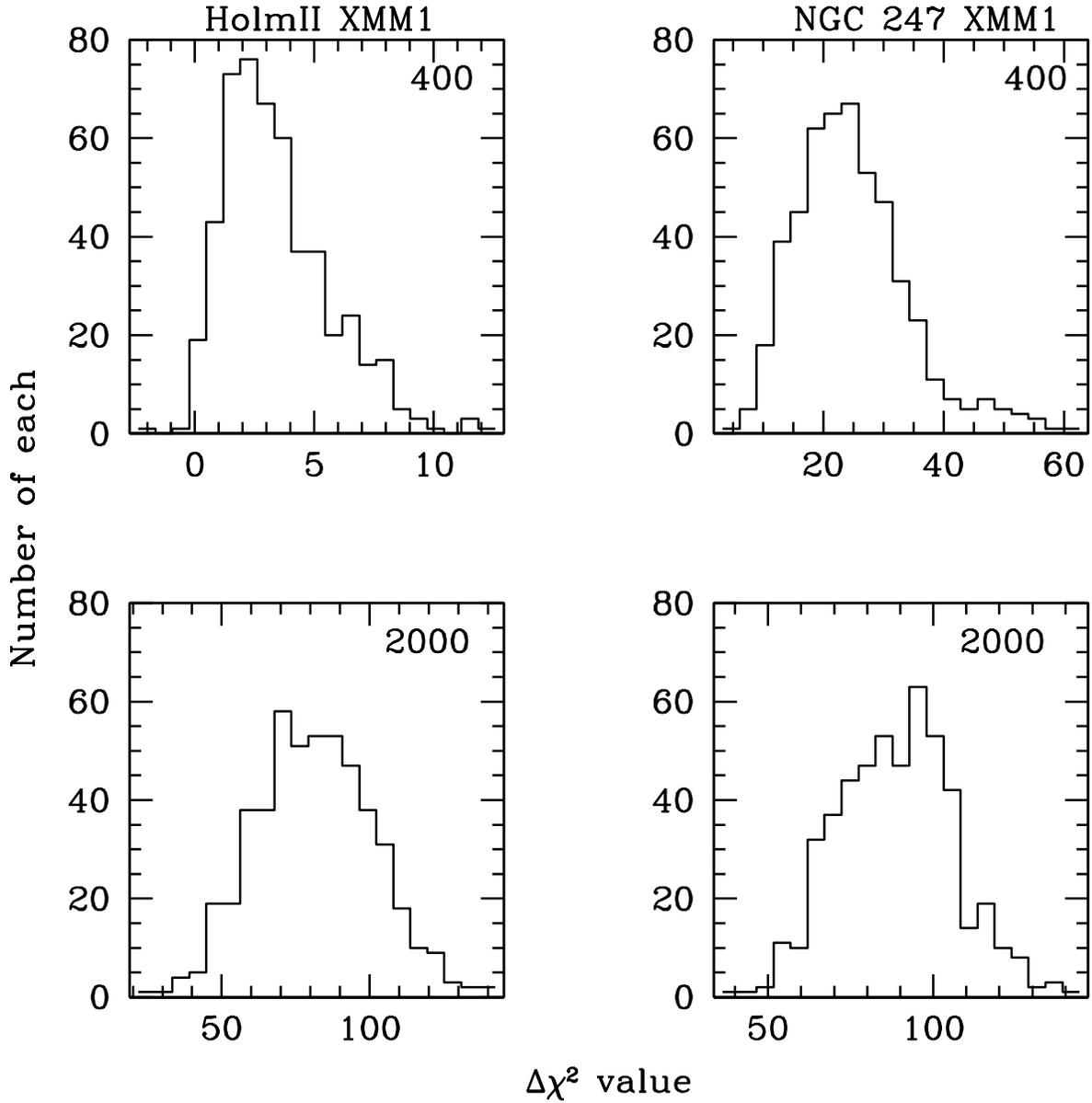}
\caption{Distribution of $\Delta\chi^2$ from simulations. The
$\Delta\chi^2$ values represent the difference between the unreduced
$\chi^2$ of the absorbed power law model and the combined blackbody
and power law model.  The left panel shows the
results for the ``weak'' blackbody component relative to power law for
500 simulated spectra at 400 counts (top) and 2000 counts (bottom).
The right panel shows the results for the ``strong'' blackbody
component.  For a weak blackbody component at 400 counts, the thermal 
component is undetectable.  However, it is able to be distinguished
as the number of counts is increased.  A strong blackbody component
is easily distinguishable at 400\,counts.
\label{fig10}}
\end{figure*}
\clearpage

\subsection{Simple Power Law Model}   
 
Our next set of simulations sought to determine our confidence in the
simple power law fit being an adequate descriptor of the spectra.
Binning provides a problem in distinguishing between a powerlaw and a
curvature in the spectrum at the low energy range, since the
binning procedure can wash out a low kT blackbody from the spectrum.
For this reason, we chose to simulate unbinned PN spectra for a source
we categorized as a low-state object, IC 0342 XMM1.  IC 0342 XMM1
represents characteristics typical of our low-state candidates, namely
it is within the proper luminosity range and it has a power law index
of $\Gamma \approx 1.7$ (the median of the distribution for low-state
objects is 2.03) and a hydrogen column density near the median of all
the fitted values (where the median value is $\approx 3 \times 10^{21}$\,cm$^{-2}$
and the value of IC 0342 X-1's column density is $5.8 \times 10^{21}$\,cm$^{-2}$).  
We chose this source for these reasons and the high
number of counts in its PN spectrum.
Instead of using the $\chi^2$ statistic (used for binned data), we
chose to use the maximum-likelihood statistic, {\it C-stat}, in {\it
XSPEC} (which uses unbinned data).  

We simulated 500 spectra using the best-fit parameters for the simple
power law fit using the {\it fakeit} command.  We fit the simulated
spectra with two models: the simple powerlaw and a combined blackbody
and power law model and computed the change in the goodness of fit,
$\Delta\chi^2$.  For the two-component model, once again we placed the
constraint that the blackbody temperature remain in the range that it
would be detectable by {\it XMM-Newton}, 0.07 - 4\,keV.  We followed
this procedure for 1000, 2000, and 4000 counts.  At both the 1000 and
2000 count level, the addition of a second component has no effect on
the C-statistic (the distributions in C-space for both the power law
fit and the two-component fit are indistinguishable).

At the 4000 count level, the C-statistic distributions for the two
models separate such that there is an 26\% confidence that the
two-component model is a better fit to the data.  We find, when we
examine the model parameters, that the power law index ($\Gamma$) for
the two-component model ranges between $1.19 - 2.12$ with the mean
value $\approx 1.74$.  The mean value for all three count levels used
clustered around this value, though the range in $\Gamma$ increased as
the counts decreased.  The mean blackbody temperature for simulations
with 4000 counts was 1.17\,keV (with the range varying between the
amount previously noted) with a median of 1.06\,keV.  For simulations
with lower counts, the blackbody temperature becomes higher (1.24\,keV
for 1000 counts) with a higher median (1.83\,keV for 1000 counts).
This tells us that the fitting procedure tends to approximate a pure
power law spectrum as a two-component spectrum with $\Gamma$ equal to
that of the true spectral index but with a blackbody temperature
higher than those observed in our study (1.1\,keV or higher) which in
the {\it XMM} band can be approximated as a power law.  If we found
spectra in our sample that were best fit with a low spectral index and
a high blackbody temperature, we might suspect that the spectrum's
true nature is a power law.  We also note that if the hydrogen
column density is large, much $> 3 \times 10^{21}$\,cm$^{-2}$, a low
temperature (kT) blackbody component can be much more difficult to
detect.

\section{Additional Spectral Fits}

The following sources were not best fit by the standard models employed in this study:

\subsection{NGC300 XMM4}
This source was classified as a super-soft X-ray source by \citet{kong03}.  We find that the standard single-component absorbed 
blackbody model is a much better model for this spectrum.  In fact, the power law, bremsstrahlung, and combined models do not 
fit the data within the 90\% confidence range.  Fitting an absorbed blackbody, we find the best fit corresponds to the following
parameters: $n_H = 1.38^{+0.27} _{-0.55}\times 10^{21}$\,cm$^{-2}$, $kT = 0.059^{+0.007} _{-0.005}$\,keV, and $\chi^2/dof = 74.5/45$.  This fit yields an
unabsorbed flux of $3.3\times10^{-13}$\,erg\,cm$^{-2}$\,s$^{-1}$. 

\subsection{NGC4631 XMM4}
The spectrum of this source clearly identifies it as a super-soft X-ray source.  As with NGC300 XMM4, the standard models
employed in this study did not adequately match the data.  The best fitting model corresponds again to an absorbed blackbody.
The corresponding parameters are as follows:  $n_H = 6.2^{+0.26} _{-1.5}\times 10^{21}$\,cm$^{-2}$, $kT = 0.07^{+0.01} _{-0.01}$\,keV, 
and $\chi^2/dof = 142.3/74$.  
This fit yields an
unabsorbed flux of $9.5\times10^{-12}$ erg\,cm$^{-2}$\,s$^{-1}$.  The position of this source shows it to be coincident
with a globular cluster associated with that galaxy.  This source was identified as a bulge X-ray source, possibly powered
by accretion, in a ROSAT study of NGC4631 \citep{vog96}.

\subsection{NGC4631 XMM5}
The spectrum of this source was best fit with an absorbed power law + an absorbed {\it vapec} model.  This indicates the prescence
of hot gas, indicating a possible thermal X-ray source.

\subsection{NGC4945 XMM5}
The spectrum of this source was not adequately fit with any of the standard models used in this investigation.  The spectrum
exhibits a prominent Fe K line in the PN spectrum that is well fit by a gaussian ({\it zgauss}) at 6.4 keV.  We find
that the entire spectrum is best fit with a partial covering fraction absorption model ({\it pcfabs}) in combination with the 
normal absorption, a power law, and a gaussian.  The best fit parameters yield: absorption column density, $n_H = 1.79\times 10^{21}$\,cm$^{-2}$, 
partial covering absorption, $n_H = 18.4\times 10^{21}$\,cm$^{-2}$, partial covering fraction $=0.82$, $\Gamma= 1.6$, and $\chi^2/dof = 61.8/57$.
The source is clearly located within the optical galaxy, and is thus unlikely to be a background AGN.   

\subsection{M51 XMM5}
The spectrum and luminosity (L$_{X} \approx 1.9 \times 10^{42}$\,erg\,s$^{-1}$) of this source suggests that it is an AGN.  
The location of the source, from the Digital Sky 
Survey, places it within the dwarf companion of M51 making a value of the optical flux hard to constrain.  
The best fit to this source was an absorbed blackbody + power law and the spectral 
parameters are listed in Table 4.

\subsection{M83 XMM2}
Like NGC4945 XMM5, this source was best fit by a partial absorption model.  However, this source showed no evidence of an
Fe K line.  We fit this source's spectra using a partial covering fraction absorption model in combination with the normal
absorption model and a power law. The best fit parameters yield: absorption column density, $n_H = 2.1 \times 10^{21}$\,cm$^{-2}$, 
partial covering absorption, $n_H = 43.5\times 10^{21}$\,cm$^{-2}$, partial covering fraction $=0.86$, $\Gamma= 2.95$, and $\chi^2/dof = 83.5/84$.  The
unabsorbed flux in the range of 0.3-10 keV equals $1.37\times 10^{-12}$\,erg\,cm$^{-2}$\,s$^{-1}$.

\subsection{Inverse Compton Scattering Sources}
Table C4 includes the parameters for the ``ULX'' sources best fit by the {\it compST} model.  A discussion of these sources and interpretation
of the data is included in section 4.3.


\section{Additional Tables}
\renewcommand{\thetable}{C\arabic{table}}

\begin{deluxetable}{llllllll}
\tabletypesize{\scriptsize}
\rotate
\tablecaption{Bright Point Sources Examined\label{tbl-2}}
\tablewidth{0pt}
\tablehead{
\colhead{Source\tablenotemark{1}} & \colhead{RA (h m s)} & \colhead{Dec ($\circ\ \prime\ \prime\prime$)} &
\colhead{Total Counts} & \colhead{Count Rate\tablenotemark{2}} &
\colhead{ID} & \colhead{Location in galaxy\tablenotemark{3}} & \colhead{XMM ref\tablenotemark{7}}
}
\startdata
NGC247 XMM1 & 00 47 03.8 & -20 47 46.2 & 3458, 1389, 1379 & 20.33, 5.8, 6.4 & 1RXS J004704.8-204743   & sa & -\\
NGC247 XMM2 & 00 47 03.1 & -20 37 02.5 & -, 597, 600      & -, 1.9, 1.4     & -            & sa & - \\
NGC253 XMM1 & 00 47 32.8 & -25 17 52.6 & -, 3156, 2985    & -, 11.38, 9.9   & NGC253 PSX-2\tablenotemark{4} & near center & 1\\
\nodata     & \nodata    & \nodata     & -, 12654, 12812  & -, 8.7, 9.1     & \nodata      & - 	& \\ 
NGC253 XMM2 & 00 47 22.4 & -25 20 55.2 & -, 825, 942      & -, 2.8, 2.97    & NGC253 PSX-5 & sa & 1\\
\nodata     & \nodata    &  \nodata    & -, 10347, 10304  & -, 8.2, 8.5     & \nodata      & - 	& \\
NGC253 XMM3 & 00 47 35.2 & -25 15 13.8 & -, 870, 1065     & -, 3, 3.41      & NGC253 PSX-7 & sa & 1\\
\nodata     & \nodata    & \nodata     & -, 5988, 6131    & -, 4.2, 4.6     & \nodata      & - 	& \\
NGC253 XMM4 & 00 47 23.3 & -25 19 06.5 & -, 649, 703      & -, 1.4, 1.3     & -            & sa & -\\
\nodata     & \nodata    & \nodata     & -, 3823,3738     & -, 1.7, 1.95    & -            & - 	& \\
NGC253 XMM5 & 00 47 17.6 & -25 18 12.1 & -, 295, 313      &  -, 1.08, 1.04  & NGC253 PSX-4 & sa & 2\\
\nodata     & \nodata    &  \nodata    & -, 4199, 4303    & -, 3.6, 3.8     & \nodata      & - 	& \\ 
NGC253 XMM6 & 00 47 42.8 & -25 15 05.5 & -, 6081, 6407    & -, 3.8, 4.4     & NGC 0253 [VP99] X40 & sa	& 1\\
NGC253 XMM7\tablenotemark{5} & 00 47 09.2 & -25 21 21.7 & -, 4300, 4454    & -, 2.6, 3.0     & -  & sa	& - \\
NGC300 XMM1 & 00 55 09.9 & -37 42 13.9 & 6778, 2248, 2453 & 18.6, 4.9, 5.4  & -            & sa 	& -\\
NGC300 XMM2 & 00 55 10.6 & -37 48 36.7 & 1364, 456, 463   & 3.1, 0.9, 0.9   & -            & edge sa? 	& -\\
NGC300 XMM3 & 00 54 49.7 & -37 38 53.8 & 915, 442, 435    & 2.2, 0.9, 0.9   & -            & sa\tablenotemark{6} & -\\
NGC300 XMM4 & 00 55 10.9 & -37 38 53.8 & 745, 224, 233    & 1.9, 0.3, 0.4   & XMMU J005511 -3749; SSS& sa & 3\\
NGC300 XMM5 & 00 55 21.1 & -37 29 19.5 & 750, 247, 250    & 1.7, 0.4, 0.5   & -            & out 	& -\\
NGC300 XMM6 & 00 54 44.2 & -37 51 04.5 & 517, 187, 165    & 1.1, 0.3, 0.3   & -            & out 	& -\\
NGC625 XMM1\tablenotemark{8} & 01 35 06.8 & -41 26 17.1 & 5832, 577, 2119  & 3, 0.6, 1.4     & - & sa 	& - \\
NGC1313 XMM1& 03 18 19.9 & -66 29 10.7 & 2876, 900, 810   & 8.6, 3.2, 2.8   & NGC1313 [CPS95] X-1; IXO 07  & sa & 4,5\\
NGC1313 XMM2& 03 17 38.8 & -66 33 05.3 & 7568, 2357, 2108 & 25.4, 8.8, 7.8  & NGC1313 [CPS95] X-3; SN 1978 K & edge sa & 6 \\
NGC1313 XMM3& 03 18 22.5 & -66 36 06.2 & 6960, 2179, 1793 & 23.3, 8.1, 6.6  & NGC1313 [CPS95] X-2; IXO 08  & edge sa? & 4,5\\
NGC1313 XMM4& 03 18 18.5 & -66 30 05   & 2075, 659, 567   & 5.1, 1.8, 1.5   & NGC1313 [SPC2000] X-8  & near center & -\\  
IC0342 XMM1 & 03 45 55.8 & +68 04 54.5 & 1802, 1216, 1105 & 33.6, 12.4, 11.1& IC0342 [RW2000] X-1; IXO 22  & sa 	& 5, 7, 8\\
IC0342 XMM2 & 03 46 15.0 & +68 11 11.2 & 1147, 541, 184   & 21.1, 5.5, 1.7  & IC0342 [RW2000] X-3    & sa 	& 7, 8\\ 
IC0342 XMM3 & 03 46 48.6 & +68 05 43.2 & 1186, 670, 606   & 21.7, 6.8, 6.0  & IC0342 [LLJ2000] X-2   & near center 	& 7, 8\\
IC0342 XMM4 & 03 46 57.2 & +68 06 20.2 & 551, 338, 377    & 9.5, 3.9, 3.6   & IC0342 [RW2000] X-6    & near center 	& 7, 8\\
NGC1569\tablenotemark{9}     & \nodata    & \nodata     & \nodata          & \nodata         & \nodata          &\nodata & \nodata\\
NGC1705 XMM1& 04 54 57.6 & -53 24 23.5 & 1174, 400, 371   & 2.4, 0.6, 0.6   & RX J0454.9-5324        & out? & -\\
NGC1705 XMM2& 04 54 19.6 & -53 20 41.9 & 933, 375, 397    & 1.9, 0.6, 0.6   & \nodata                & out? & -\\
NGC1705 XMM3& 04 54 38.1 & -53 18 16.2 & 698, 372, 418    & 1.4, 0.6, 0.6   & WGA J0454.7-5318       & out? & -\\
MRK71 XMM1  & 07 28 51.8 & +69 07 27   & 832, 225, 207    & 3.9, 0.97, 0.9  & -             & out 	& -\\
NGC2403 XMM1& 07 36 25.6 & +65 35 40   & -, 1199, 672     & -, 1.0, 0.60    & NGC2403 [RW2000] X-1       & edge sa & - \\
NGC2403 XMM2& 07 36 50.2 & +65 36 02.1 & 1964, 729, 672   & 1.99, 0.63, 0.60& -              & near center & -\\
NGC2403 XMM3& 07 36 55.4 & +65 35 40.3 & 1497, 489, 378   & 1.4, 0.40, 0.33 & -              & near center & -\\
NGC2403 XMM4& 07 37 02.5 & +65 39 35.2 & 1004, 274, 288   & 0.52, 0.15, 0.21& NGC2403 [RW2000] X-4 & edge sa? & -\\
Hol II XMM1 & 08 19 28.8 & +70 42 20.3 & 31052, 1257, 10807& 272.2, 75.5, 72.7 & Holm II X-1; IXO 31\tablenotemark{10} & near center? & 9\\
\nodata     & \nodata    & \nodata     & 3853, 1361, 1452 & 78.3, 18.9, 20.5 & \nodata       & -  & \\
Hol I XMM1  & 09 41 30   & +71 12 34   & 687, 768, 754    & 2.9, 2.7, 2.6   & -              & out? 	& -\\
Hol I XMM2  & 09 39 59.7 & +71 06 40.2 & 575, 203, 224    & 2.5, 0.7, 0.7   &  1WGA J0940.0+7106            & out? & -\\
Hol I XMM3  & 09 42 06.7 & +71 04 45.3 & 452, 141, 141    & 1.7, 0.4, 0.4   & -              & out? & -\\
M81 XMM1    & 09 55 32.9 & +69 00 34.8 & 50788, -, 18988  & 51.1, -, 21.7   & NGC3031 [RW2000] X-11  & sa & -\\
\nodata     & \nodata    & \nodata     & -, 1227, -\tablenotemark{11}     & -, 12, 13.1     & \nodata        & - & -\\
M81 XMM2    & 09 55 24.8 & +69 01 11.7 & 17871, -, 4121   & 13.2, -, 4.4    & SN 1993J       & sa & 10\\
M81 XMM3    & 09 55 10.6 & +69 05 02.2 & -, -, 1970       & -, -, 1.5       & NGC3031 [RW2000] X-05 & sa & -\\
M81 XMM4    & 09 55 24.3 & +69 10 00.2 & -, -, 1197       & -, -, 1.0       & NGC3031 [RW2000] X-08   & edge sa & -\\
M81 XMM5    & 09 55 49.2 & +69 05 30.5 & -, -, 2077       & -, -, 2.3       & -              & sa  & -\\
M82\tablenotemark{12}&\nodata&\nodata   & \nodata          &\nodata          &\nodata         &\nodata  & \nodata\\
Hol IX XMM1 & 09 57 53.3 & +69 03 48.7 & 14976, 6546, 6586& 207.3, 64.3, 65 & Hol IX X-1; IXO 34   & ? & 4, 5\\
Sextans A XMM1& 10 11 24.6& -04 42 17.2& 3963, 1323, 1242 & 17.2, 5, 4.5    & -              & out & - \\
IC2574 XMM1 & 10 28 42.4 & +68 28 17.8 & 1047, 673, 623   & 8.3, 2.6, 2.5   & -              & sa & -\\
IC2574 XMM2 & 10 26 33.5 & +68 29 32.1 & 533, 335, 300    & 4.3, 1.3, 1.2   & -              & out & -\\
IC2574 XMM3 & 10 27 22.2 & +68 18 47.6 & 538, 293, 301    & 4.2, 1.1, 1.2   & -              & out & -\\
NGC4214 XMM1& 12 15 37.0 & +36 19 29.4 & 434, 230, 225    & 2.9, 1.6, 1.5   & NGC4214 [HSS2004] 11              & sa & -\\
NGC4214 XMM2& 12 15 58.2 & +36 22 38.5 & 626, 160, 285    & 3.4, 0.6, 1.5   & -              & out & -\\
NGC4258 XMM1\tablenotemark{5}& 12 18 47.8 & +47 20 51.7 & 828, 470, 444    & 7.0, 2.8, 2.8   & -              & sa & -\\
\nodata     & \nodata    & \nodata     & \nodata          & \nodata         & \nodata        & - & -\\
NGC4258 XMM2& 12 18 57.8 & +47 16 06.8 & 732, 337, 337    & 5.8, 2, 2.1     & NGC4258 [RW2000] X-7 & sa & -\\
\nodata     & \nodata    & \nodata     & 716, 268, 290    & 1.8, 1, 1.0     & \nodata        & - & \\
NGC4258 XMM3& 12 18 56.5 & +47 21 24.3 & 489, 182, 184    & 3.8, 0.9, 1.1   & NGC4258 [RW2000] X-5 & sa & -\\
\nodata     & \nodata    & \nodata     & -, 160, 142      & -, 0.3, 0.3     & \nodata        & - & \\
NGC4258 XMM4& 12 19 23.2 & +47 09 37.2 & 644, 277, 224    & 4.9, 1.3, 1.3   & HELLAS 288     & edge sa? & -\\
\nodata     & \nodata    & \nodata     & 964, 410, 382    & 0.3, 0.2, 0.1   & \nodata        & - & \\
NGC4395 XMM1& 12 26 01.5 & +33 31 29   & 2162, 862, 921   & 15.5, 5.1, 5.6  & NGC4395 [RW2000] X-1; IXO 53 & sa & -\\
NGC4395 XMM2& 12 25 25.3 & +33 36 46.4 & 392, 252, 208    & 2.6, 1.5, 1.2   & -              & sa? & -\\
NGC4395 XMM3& 12 25 32.6 & +33 25 27.9 & 763, 277, 278    & 5.2, 1.5, 1.6   & -              & out? & -\\
NGC4395 XMM4& 12 25 42.7 & +33 40 00.1 & 516, 60, 40      & 3.1, 0.3, 0.2   & -              & out & -\\
NGC4449 XMM1& 12 28 18   & +44 06 30.9 & 1409, 608, 593   & 10.6, 3.6, 3.5  & XRB?; NGC4449 [RW2000] X-7; Source 27\tablenotemark{13}& sa & -\\
NGC4449 XMM2& 12 28 09.3 & +44 05 03.9 & 1503, 527, 586   & 11.3, 3.1, 3.5  & SNR?; NGC4449 [RW2000] X-1; Source 10 & sa & -\\
NGC4449 XMM3& 12 28 11.1 & +44 06 43.9 & 1094, 549, 404   & 8.3, 3.2, 2.3   & SNR; Source 15           & sa & -\\
NGC4490 XMM1& 12 30 32.4 & +41 39 14.6 & 746, 323, 393    & 6.0, 1.9, 2.2   & NGC4490 [RW2000] X-1 & sa & -\\
NGC4490 XMM2& 12 30 36.5 & +41 38 33.3 & 656, 299, 310    & 5.3, 1.7, 1.7   & NGC4490 [RW2000] X-2 & near center & -\\
NGC4490 XMM3& 12 30 43.3 & +41 38 11.5 & 832, 501, 461    & 5.8, 2.6, 2.2   & NGC4490 [RW2000] X-4 & sa & -\\
NGC4490 XMM4& 12 30 31.1 & +41 39 08.1 & 546, 291, 286    & 4.4, 1.7, 1.6   & CXOU J123030.8 +413911 & sa & -\\
NGC4490 XMM5& 12 30 30.3 & +41 41 40.3 & 413, 587, 482    & 2.7, 3.1, 2.3   & NGC4485 [RW2000] X-1 & sa & -\\
NGC4631 XMM1& 12 41 55.8 & +32 32 14   & 5093, 1969, 1762 & 13, 4.2, 3.8    & NGC4631 [RW2000] X-1; IXO 68 & sa & -\\
NGC4631 XMM2& 12 41 57.5 & +32 32 01   & 1273, 531, 400   & 3.3, 1.2, 0.9   & NGC4631 [RW2000] X-2 & sa & - \\
NGC4631 XMM3& 12 41 58.2 & +32 28 49.6 & 1271, 443, 457   & 2.8, 0.9, 0.9   & CXOUSEXSI J124158.0+322851 & out & -\\
NGC4631 XMM4& 12 42 16.1 & +32 32 48.8 & 957, 400, 429    & 2.3, 0.8, 0.8   & [VP96] H13 & sa & -\\
NGC4631 XMM5& 12 42 11.2 & +32 32 33.6 & 1626, 894, 866   & 3.9, 1.9, 1.8   & [VP96] H12; [HFE2003] PSX-01 & sa & -\\
NGC4736 XMM1& 12 50 50.2 & +41 07 12   & 713, 273, 227    & 6.7, 2.1, 1.7   & NGC4736 X-4\tablenotemark{14}& near center & -\\
NGC4945 XMM1& 13 05 33.3 & -49 27 36.3 & 1456, 646, 595   & 7.9, 2.9, 2.6   & NGC4945 [GMB2000] X-2& sa & -\\
NGC4945 XMM2& 13 05 38.4 & -49 25 45.3 & 1393, 600, 523   & 7.6, 2.7, 2.3   & NGC4945 [R97] X-3 & sa & -\\
NGC4945 XMM3& 13 05 18.8 & -49 28 24   & -, 357, 362      & -, 1.6, 1.6     & \nodata & sa & -\\
NGC4945 XMM4& 13 05 22.2 & -49 28 26.3 & 731, 331, 332    & 4.0, 1.5, 1.5   & NGC4945 [BIR96] X-1? & sa & -\\
NGC4945 XMM5& 13 05 25.7 & -49 28 30.7 & 772, 267, 301    & 4.2, 1.2, 1.32  & \nodata  & sa & -\\
NGC5204 XMM1& 13 29 38.5 & +58 25 03.6 & 9981, 3352, 3384 & 62.8, 17.7, 17.9& NGC5204 [RW2000] X-1; IXO 77 & sa & 11\\
\nodata     & \nodata    & \nodata     & 9231, 2284, 2349 & 85.5, 24.7, 25.8& \nodata        & -  & \\
NGC5204 XMM2& 13 29 27.4 & +58 25 31.8 & 573, 170, 231    & 3.4, 0.8, 1.2   & -              & edge sa & -\\
\nodata     & \nodata    & \nodata     & 772, 161, 121    & 5.0, 1.7, 1.1   & \nodata        & - & \\
M51 XMM1    & 13 29 40   & +47 12 36.2 & 1102, 367, 409   & 6.2, 1.8, 2     & NGC5194 [RW2000] X-1              & sa  & -\\
M51 XMM2    & 13 30 07.7 & +47 11 04.8 & 514, 540, 549    & 2.6, 2.2, 2.2   & IXO 81              & sa & -\\
M51 XMM3    & 13 30 01.1 & +47 13 41.4 & 1004, 315, 311   & 4.1, 1.1, 1.1   & CXOU J133001.0 +471344; IXO 80              & sa & -\\
M51 XMM4    & 13 30 06   & +47 15 38.9 & 518, 183, 166    & 2.8, 0.9, 0.8   & NGC 5195 [RW2000] X-1              & sa & -\\
M51 XMM5    & 13 29 59.6 & +47 15 54   & 1079, 359, 257   & 5, 1.5, 1       & CXOU J132959.5 +471559              & near center & -\\
M51 XMM6    & 13 29 57.5 & +47 10 45.3 & 536, 206, 247    & 1.9, 0.6, 0.8   & CXOU J132957.6 +471048              & sa & -\\
M51 XMM7    & 13 29 53.6 & +47 14 31.5 & 452, 141, 143    & 2.4, 0.6, 0.7   & CXOU J132953.8 +471432              & edge sa & -\\
M83 XMM1    & 13 37 19.8 & -29 53 49.8 & 3074, 927, 987   & 12, 3.3, 2.5    & RX J133719 -2953.6; IXO 82 & sa & -\\
M83 XMM2    & 13 36 59.4 & -29 49 57.2 & 1133, 371, 397   & 4, 1.3, 1       & CXOU J133659.4 -294959 & sa & -\\
M83 XMM3\tablenotemark{15}    & 13 37 04.4 & -29 51 24   & 1724, 576, 459   & 7.2, 2.3, 1.2   & CXOU J133704.3 -295121 & sa & -\\
M83 XMM4    & 13 37 01.5 & -29 53 26   & 1289, 345, 401   & 4.9, 1.3, 1.0   & CXOU J133701.4 -295326 & sa & -\\
NGC5253\tablenotemark{16}     & \nodata    &\nodata       &\nodata           &\nodata         & \nodata        & \nodata & \nodata \\
M101 XMM1   & 14 03 14.7 & +54 18 05   & 2690, 1449, 1417 & 10.3, 3.3, 3.3  & CXOU J140313.9 +541811; XMM-2\tablenotemark{17} & sa & 12\\
M101 XMM2   & 14 03 03.8 & +54 27 37   & 2825, 1623, 1551 & 10.3, 3.6, 3.3  & XMM-1          & edge sa? & 12\\
M101 XMM3   & 14 04 14.6 & +54 26 04.4 & 1460, 822, 717   & 5, 1.6, 1.3     & CXOU J140414.3 +542604; XMM-3 & edge sa? & 12\\
M101 XMM4   & 14 02 28.5 & +54 16 26.7 & 1505, 877, 757   & 5.1, 1.7, 1.5   & CXOU J140228.3 +541626 & sa & 12\\
M101 XMM5   & 14 02 22.5 & +54 17 58   & 516, 245, 289    & 1.8, 0.4, 0.6   & CXOU J140222.2 +541756; XMM-6 & sa & 12\\
NGC5408 XMM1& 14 03 19.8 & -41 22 59.3 & 5932, 2036, 2077 & 12.8, 3.2, 3.3  & NGC5408 [KCP2003] X-1 & sa & 5\\
CIRCINUS XMM1\tablenotemark{18}& 14 12 54.2& -65 22 55.3 & 16220, 11452, -  & 14.5, 11.3, -   & -              & edge sa? & -\\
CIRCINUS XMM2& 14 12 39.2& -65 23 34.3 & 8741, 2386, -    & 5.7, 1.9, -     & -              & edge sa? & -\\
CIRCINUS XMM3& 14 13 28.3& -65 18 08.3 & 4873, 1031, -    & 1.8, 0.7, -     & -              & edge sa? & -\\
\enddata
\tablenotetext{1}{Sources labeled {\it XMM-n} in order of apparent brightness from the first observation studied}
\tablenotetext{2}{count rate units of $\times 10^{-2}$\,cts\,s$^{-1}$ for the PN, MOS1, and MOS2 spectra}
\tablenotetext{3}{location specified as: inside the optical extent of the galaxy or in spiral arms (sa), 
near the center of the galaxy, at the edge of a spiral arm/ optical extent of galaxy, or outside the optical extent 
of the galaxy.  Location based on DSS images.}
\tablenotetext{4}{identification for NGC0253 following Humphrey et al. (2003)}
\tablenotetext{5}{Transient.}
\tablenotetext{6}{Appears as an extended source in HST image.}
\tablenotetext{7}{references to studies using {\it XMM-Newton} spectra}
\tablenotetext{8}{Spectra too scattered to model.}
\tablenotetext{9}{bright sources coincide with nucleus (unresolvable star cluster and X-ray binaries), fore-ground star, and background AGN (Martin, Kobulnicky, Heckman (2002)}
\tablenotetext{10}{IXO designation from Colbert \& Ptak (2002)}
\tablenotetext{11}{Spectrum from Hol IX observation.}
\tablenotetext{12}{bright source is too close to other sources}
\tablenotetext{13}{Chandra observations of point sources in NGC4449 published in Summers et al. (2003); Source 27 varied from ROSAT observations}
\tablenotetext{14}{three other sources near the nucleus are unresolveable in the XMM obs., but seen by Chandra (Eracleous et al. 2002)}
\tablenotetext{15}{Unable to model sprectrum due to an error in $\chi^2$-space}
\tablenotetext{16}{bright sources are too close, but resolvable by Chandra (Summers et al. 2004)}
\tablenotetext{17}{alternate ID from Jenkins et al. (2004)}
\tablenotetext{18}{bright sources near the nucleus are unresolvable, but seen by Chandra (Smith \& Wilson 2001)}

\tablerefs{
(1) Pietsch et al. 2001; (2) Pietsch, Haberl, \& Vogler 2003; (3) Kong \& Di Stefano 2003;
(4) Miller et al. 2004;	 (5) Wang et al. 2004; (6) Schlegel et al. 2004;
(7) Bauer, Brandt, \& Lehmer 2003; (8) Kong 2003; (9) Lehmann et al. 2005;
(10) Zimmermann \& Aschenbach 2003; (11) Roberts et al. 2005; (12) Jenkins et al. 2004.}
\end{deluxetable}

\clearpage

\begin{deluxetable}{llll}
\tabletypesize{\scriptsize}
\tablecaption{Bright, Identifiable Background and Foreground Sources\label{tbl-7}}
\tablewidth{0pt}
\tablehead{
\colhead{Galaxy} & \colhead{RA (h m s)} & \colhead{Dec ($\circ\ \prime\ \prime\prime$)} & \colhead{Identification}
}

\startdata
NGC 247  & 0 46 51.7  & -20 43 30   & QSO B044-2059 \\
NGC 300  & 0 55 26.7  & -37 31 25.6 & HD 5403 (Star)\\
NGC 625  & 01 34 42.4 & -41 36 15.2 & QSO B0132-4151 \\
NGC 1569\tablenotemark{a} & 04 31 16.9 & +64 49 50   & CXOU J043116.8+644950 (Star) \\
NGC 1569 & 04 31 14.2 & +64 51 07.9  & CXOU 043114.0+645107 (Star) \\
NGC 1569 & 04 31 25.4 & +64 51 53.8 & CXOU 043125.1+645154 (AGN) \\
NGC 1705 & 04 54 01.2 & -53 21 12.3 & WGA J0454.0-5320 (M star or elliptical galaxy) \\
NGC 2403 & 07 35 09   & +65 40 27.5 & HD 59581 (Star)\\
NGC 4258 & 12 18 08.9 & +47 16 08.3 & QSO J1218+472 \\
M83      & 13 36 45.6 & -29 59 13.9 & 2MASX J13364579-2959122 (Galaxy) \\
M83      & 13 36 13.9 & -29 56 13   & RX J133615-2957.8 (Galaxy) \\
NGC 5253 & 13 39 50.6 & -31 34 11.1 & CD-30 10790 (Star) \\
M101     & 14 02 30   & +54 21 18.2 & [WIP99] H13 (Star)\tablenotemark{b} \\
NGC 5408 & 14 03 27.5 & -41 25 18.5 & (Star) \\
\enddata

\tablenotetext{a}{identification for objects in NGC 1569 from Martin, Kobulnicky, \& Heckman (2002)}
\tablenotetext{b}{confirmed by K. Kuntz using HST ACS}
\end{deluxetable}

\clearpage

\begin{deluxetable}{llllll}
\tabletypesize{\scriptsize}
\tablecaption{{\it XMM-Newton} power law fit for best fit two-component spectra\label{tbl-6}}
\tablewidth{0pt}
\tablehead{
\colhead{Source} & \colhead{n$_H$\tablenotemark{a}} & \colhead{$\Gamma$} & \colhead{$\chi^{2}/$dof} &
\colhead{$F_X$\tablenotemark{c}} & \colhead{$L_X$\tablenotemark{d}}
}

\startdata
NGC247 XMM1   & $9.5^{+1.6} _{-1.4}$ & $8.52^{+1.14} _{-0.91}$ & 112.2/95 & 1900 & 2200 \\
NGC253 XMM1   & $3.4^{+0.3} _{-0.3}$ & $1.77^{+0.06} _{-0.06}$ & 262.6/232 & 3.1 & 3.6 \\
\nodata       & $6.9^{+0.4} _{-0.4}$ & $1.98^{+0.05} _{-0.05}$ & 611.6/582 & 2.9 & 3.3 \\
NGC253 XMM2 (obs 2)   & $2.2^{+0.1} _{-0.1}$ & $2.03^{+0.04} _{-0.04}$ & 507.4/500 & 1.2 & 1.4 \\
NGC253 XMM3   & $3.9^{+0.5} _{-0.4}$ & $2.17^{+0.14} _{-0.11}$ & 91.9/83 & 0.73 & 1.2 \\ 
\nodata       & $4.0^{+0.4} _{-0.3}$ & $2.06^{+0.09} _{-0.07}$ & 381.8/409 & 0.98 & 1.6 \\ 
NGC253 XMM4   & $8.5^{+3.0} _{-2.3}$ & $2.09^{+0.33} _{-0.28}$ & 73.6/59 & 0.52 & 0.85 \\
\nodata       & $1.2^{+0.3} _{-0.3}$ & $2.09^{+0.16} _{-0.15}$ & 321.4/293 & 0.29 & 0.48 \\
NGC253 XMM5   & $1.7^{+1.2} _{-0.9}$ & $1.54^{+0.22} _{-0.20}$ & 31.8/25 & 0.32 & 0.53 \\
\nodata       & $3.4^{+0.2} _{-0.2}$ & $2.17^{+0.7} _{-0.7}$ & 283.8/298 & 1.1 & 1.3 \\
NGC253 XMM6   & $3.9^{+0.3} _{-0.3}$ & $2.21^{+0.84} _{-0.80}$ & 435/409 & 0.93 & 1.5 \\
NGC253 XMM7   & $7.1^{+0.7} _{-0.7}$ & $2.15^{+0.11} _{-0.11}$ & 357/342 & 1.3 & 2.2 \\ 
NGC300 XMM1   & $0.97^{+0.11} _{-0.11}$ & $2.67^{+0.06} _{-0.06}$ & 469.8/422 & 0.81 & 0.6 \\ 
NGC300 XMM2   & $1.7^{+0.40} _{-0.40}$ & $3.20^{+0.32} _{-0.24}$ & 133.98/99 & 0.27 & 0.21 \\
NGC300 XMM3   & $3.4^{+0.8} _{-0.6}$ & $1.86^{+0.15} _{-0.13}$ & 101.9/81 & 0.19 & 0.15 \\
NGC300 XMM5   & $0.311$\tablenotemark{e} & $2.29^{+0.15} _{-0.14}$ & 54.2/56 & 0.14 & 0.10 \\
NGC300 XMM6   & $?^{+?} _{-?}$ & $2.05^{+0.17} _{-0.15}$ & 47.6/37 & 0.06 & 0.15 \\ 
NGC1313 XMM1  & $1.5^{+0.2} _{-0.2}$ & $1.81^{+0.08} _{-0.09}$ & 219.8/203 & 0.42 & 0.88 \\
NGC1313 XMM2  & $2.8^{+0.16} _{-0.16}$ & $2.48^{+0.07} _{-0.06}$ & 464.1/421 & 1.9 & 4.0 \\
NGC1313 XMM3  & $3.6^{+0.2} _{-0.2}$ & $3.2^{+0.09} _{-0.09}$ & 778.3/426 & 3.3 & 6.9 \\
IC0342 XMM3   & $3.8^{+0.4} _{-0.4}$ & $2.58^{+0.15} _{-0.14}$ & 185.8/109 & 1.7 & 3.1 \\
NGC1705 XMM1  & 0.3\tablenotemark{e} & $1.93^{+0.11} _{-0.10}$ & 61.9/88 & 0.12 & 0.37 \\
NGC1705 XMM2  & $1.4^{+0.45} _{-0.41}$ & $2.12^{+0.26} _{-0.15}$ & 91/76 & 0.078 & 0.24 \\ 
NGC1705 XMM3  & $0.6^{+0.36} _{-0.40}$ & $1.36^{+0.12} _{-0.13}$ & 80.9/67 & 0.17 & 0.53 \\ 
NGC2403 XMM1  & $3.2^{+0.61} _{-0.55}$ & $2.15^{+0.16} _{-0.15}$ & 92.2/81 & 2.2 & 3.6 \\
NGC2403 XMM2  & $2.7^{+0.37} _{-0.34}$ & $2.07^{+0.11} _{-0.11}$ & 179.5/151 & 1.3 & 2.0 \\
NGC2403 XMM3  & $1.9^{+0.40} _{-0.36}$ & $1.97^{+0.14} _{-0.13}$ & 92.6/107 & 0.81 & 1.3 \\
HolmII XMM1 (obs 1)  & $1.5^{+0.07} _{-0.07}$ & $2.61^{+0.04} _{-0.04}$ & 1134.2/976 & 12 & 10 \\
Holm I XMM1   & $?^{+?} _{-?}$ & $2.04^{+0.14} _{-0.07}$ & 102.8/95 & 0.48 & 1.7 \\
M81 XMM1      & $3.2^{+0.07} _{-0.07}$ & $2.09^{+0.02} _{-0.02}$ & 1849.9/1245 & 4.5 & 7.0 \\
\nodata       & $3.0^{+0.3} _{-0.3}$ & $1.79^{+0.07} _{-0.07}$ & 224.9/208 & 4.3 & 6.7 \\
M81 XMM2      & 7.3 & 6.13 & 1358.2/618 & 48.5 & 75.2 \\
M81 XMM3      & $0.97^{+0.25} _{-0.41}$ & $1.58^{+0.18} _{-0.15}$ & 81.35/80 & 2.5 & 3.9 \\ 
M81 XMM4      & $?^{+?} _{-?}$ & $0.88^{+0.11} _{-0.11}$ & 66.4/52 & 0.35 & 0.54 \\ 
M81 XMM5      & $1.0^{+0.4} _{-0.3}$ & $1.52^{+0.11} _{-0.11}$ & 97.5/82 & 0.44 & 0.68 \\
Holm IX XMM1  & $1.7^{+0.08} _{-0.08}$ & $1.84^{+0.03} _{-0.03}$ & 1000.9/882 & 9.4 & 15 \\
Sextans A XMM1& $0.18^{+0.23} _{-0.16}$ & $2.25^{+0.12} _{-0.07}$ & 271.4/275 & 0.56 & 0.13 \\
NGC4214 XMM2  & $0.2^{+0.5} _{-0.2}$ & $2.03^{+0.43} _{-0.28}$ & 50.9/46 & 0.16 & 0.14 \\
NGC4258 XMM1  & $1.6^{+0.4} _{-0.4}$ & $1.9^{+0.14} _{-0.13}$ & 101.4/78 & 0.06 & 0.04 \\
NGC4258 XMM2 (obs 1)  & $3.5^{+0.9} _{-0.7}$ & $1.88^{+0.16} _{-0.15}$ & 97.5/63 & 0.43 & 2.6 \\
NGC4395 XMM1  & $3.7^{+0.5} _{-0.4}$ & $4.93^{+0.34} _{-0.30}$ & 195.1/156 & 7.2 & 14 \\
NGC4395 XMM3  & $?^{+?} _{-?}$ & $1.86^{+0.14} _{-0.09}$ & 55.9/58 & 0.25 & 0.48 \\
NGC4449 XMM1  & $6.3^{+0.9} _{-0.7}$  & $2.22^{+0.14} _{-0.12}$ & 103/118 & 1.2 & 1.36 \\
NGC4449 XMM3  & $3.3^{+0.5} _{-0.4}$ & $3.36^{+0.29} _{-0.23}$ & 154/89 & 1.3 & 1.5 \\
NGC4490 XMM1  & $0.83^{+0.14} _{-0.12}$ & $2.53^{+0.17} _{-0.16}$ & 101.5/65 & 1.2 & 8.7 \\
NGC4490 XMM2  & $6.3^{+1.3} _{-1.0}$ & $2.36^{+0.19} _{-0.17}$ & 49.5/56 & 0.92 & 6.7 \\
NGC4490 XMM3  & $9.4^{+1.5} _{-1.2}$ & $2.95^{+0.24} _{-0.20}$ & 76.7/80 & 1.5 & 11 \\
NGC4631 XMM1  & $2.3^{+0.16} _{-0.15}$ & $2.13^{+0.06} _{-0.05}$ & 383.4/347 & 0.76 & 5.1 \\ 
NGC4631 XMM2  & $1.9^{+0.4} _{-0.3}$ & $2.01^{+0.14} _{-0.12}$ & 119.5/99 & 0.23 & 1.5 \\
NGC4631 XMM3  & $0.63^{+0.3} _{-0.2}$ & $1.53^{+0.1} _{-0.08}$ & 146/98 & 0.15 & 1.0 \\
NGC4736 XMM1  & $0.95^{+0.6} _{-0.5}$ & $2.02^{+0.26} _{-0.25}$ & 62.8/53 & 0.36 & 0.80 \\
NGC4945 XMM1  & $5.8^{+0.8} _{-0.7}$ & $1.88^{+0.08} _{-0.10}$ & 116/122 & 0.9 & 1.0 \\ 
NGC4945 XMM2  & $3.4^{+0.6} _{-0.5}$ & $1.58^{+0.09} _{-0.10}$ & 114.5/115 & 0.71 & 0.82 \\ 
NGC4945 XMM4  & $5.2^{+0.8} _{-0.7}$ & $2.59^{+0.19} _{-0.17}$ & 75.5/62 & 0.49 & 0.56 \\ 
NGC5204 XMM1  & $0.61^{+0.1} _{-0.1}$ & $2.11^{+0.04} _{-0.04}$ & 592.1/561 & 2.0 & 5.5 \\
\nodata       & $1.1^{+0.1} _{-0.1}$ & $2.41^{+0.07} _{-0.07}$ & 533/498  & 3.0 & 8.3 \\
M51 XMM1      & $1.1^{+0.30} _{-0.27}$ & $2.67^{+0.20} _{-0.16}$ & 110.5/82 & 0.34 & 2.8 \\
M51 XMM2      & $2.3^{+0.50} _{-0.30}$ & $2.50^{+0.22} _{-0.20}$ & 75.2/70  & 0.52 & 3.3 \\
M51 XMM5      & 2.7 & 3.08 &  256.0/72   & 0.43 & 2.7 \\
M51 XMM6      & $2.0^{+0.83} _{-0.72}$ & $2.50^{+0.33} _{-0.25}$ & 40.97/43 & 0.13 & 0.83 \\
M51 XMM7      & $0.5^{+0.39} _{-0.46}$ & $1.95^{+0.23} _{-0.18}$ & 37.8/31 & 0.11 & 0.66 \\
M83 XMM1      & $1.9^{+0.34} _{-0.31}$ & $2.32^{+0.13} _{-0.12}$ & 210.9/211 & 0.64 & 2.8 \\
M83 XMM4      & $6.0^{+1.8} _{-1.3}$ & $2.54^{+0.26} _{-0.23}$ & 91.4/91 & 0.4 & 1.8 \\
M101 XMM1     & $0.56^{+0.15} _{-0.14}$ & $1.98^{+0.08} _{-0.08}$ & 303/233 & 0.45 & 2.9 \\
M101 XMM2     & $2.2^{+0.25} _{-0.23}$ & $1.85^{+0.07} _{-0.07}$ & 288.8/263 & 0.81 & 5.3 \\ 
M101 XMM3     & $1.5^{+0.4} _{-0.3}$ & $2.70^{+0.21} _{-0.17}$ & 148.9/133 & 0.51 & 3.4 \\
M101 XMM4     & $2.2^{+0.45} _{-0.42}$ & $2.25^{+0.20} _{-0.17}$ & 165.7/140 & 0.38 & 2.5 \\ 
M101 XMM5     & $1.3^{+0.2} _{-0.3}$ & $2.28^{+0.12} _{-0.11}$ & 47.9/46 & 0.12 & 0.8 \\
NGC5408 XMM1  & $1.6^{+0.2} _{-0.1}$ & $3.57^{+0.12} _{-0.11}$ & 396.8/339 & 7.04 & 19.4 \\  
CIRCINUS XMM1 & $7.6^{+0.3} _{-0.3}$ & $2.15^{+0.05} _{-0.04}$ & 762.9/863 & 4.6 & 8.8 \\
CIRCINUS XMM2 & $11.7^{+0.4} _{-0.7}$ & $3.48^{+0.13} _{-0.06}$ & 517.9/432 & 2.7 & 5.2 \\
CIRCINUS XMM3 & $9.0^{+2.1} _{-1.0}$ & $2.57^{+0.40} _{-0.17}$ & 285.2/262 & 0.32 & 0.61 \\
\enddata

\tablenotetext{a}{total column density in units of $10^{21}$ cm$^{-2}$}
\tablenotetext{b}{improvement in $\chi^2$ over the single-component power law model}
\tablenotetext{c}{unabsorbed flux in the 0.3-10 keV band in units of $10^{-12}$ erg cm$^{-2}$ s$^{-1}$}
\tablenotetext{d}{unabsorbed luminosity in the 0.3-10 keV band, using the distances quoted in Table~\ref{tbl-1}, in units of $10^{39}$ erg s$^{-1}$}
\tablenotetext{e}{absorption level frozen at Galactic level}
\end{deluxetable}

\clearpage

\begin{deluxetable}{llllll}
\tabletypesize{\scriptsize}
\tablecaption{Best-Fit Absorbed Comptonization Model Parameters\label{tbl-9}}
\tablewidth{0pt}
\tablehead{
\colhead{ID} & \colhead{nH\tablenotemark{a}} & \colhead{kT\tablenotemark{b}} & \colhead{tau\tablenotemark{c}} 
& \colhead{$\chi^2$} & \colhead{$F_X$\tablenotemark{d}}
}
\startdata
NGC 253 XMM2 & $1.8^{+0.08} _{-0.16}$& $1.28^{+0.13} _{-0.12}$ & $19.59^{+2.0} _{-1.6}$ & 464/498 & 1.47 \\
NGC 2403 XMM1& $1.95^{+1.2} _{-0.6}$ & $0.98^{+0.16} _{-0.15}$ & $25.4^{+7.2} _{-8.3}$ & 82.8/85 & 1.4 \\
NGC 4490 XMM1& $4.7^{+1.1} _{-2.1}$ & $0.96^{+0.13} _{-0.16}$ & $27.0^{+18.3} _{-4.9}$ & 66.5/64 & 0.66 \\
NGC 4490 XMM2& $5.0^{+1.6} _{-1.4}$ & $1.21^{+0.21} _{-0.31}$ & $18.8^{+8.6} _{-9.7}$ & 45.7/55 & 0.67 \\
M101 XMM2    & $1.6^{+0.26} _{-0.24}$ & $1.24^{+0.09} _{-0.15}$ & $23.3^{+3.4} _{-2.8}$ & 256/262 & 0.65 \\
M101 XMM3    & $1.1^{+0.43} _{-0.40}$ & $1.13$ & $15.2$ & 128/132 & 0.41 \\
Circinus XMM2& $6.8^{+1.3} _{-0.9}$ & $0.62^{+0.08} _{-0.04}$ & $29.7^{+8.1} _{-7.0}$ & 437.2/430 & 0.5 \\
Circinus XMM3& $6.7^{+1.6} _{-2.4}$ & $0.93^{+0.28} _{-0.24}$ & $23.1^{+17.5} _{-5.2}$ & 273.1/261 & 0.17 \\
\enddata
\tablenotetext{a}{total column density in units of $10^{21}$ cm$^{-2}$}
\tablenotetext{b}{temperature in keV}
\tablenotetext{c}{optical depth}
\tablenotetext{d}{unabsorbed flux in the 0.3-10 keV band in units of $10^{-12}$ erg cm$^{-2}$ s$^{-1}$}
\end{deluxetable}

\begin{deluxetable}{llllll}
\tabletypesize{\scriptsize}
\tablecaption{{\it XMM-Newton} Galaxy Observations\label{tbl-10}}
\tablewidth{0pt}
\tablehead{
\colhead{Galaxy} & \colhead{S$_{60}$ (Jy)} & \colhead{S$_{100}$ (Jy)} & \colhead{F$_{FIR}$\tablenotemark{a} } &
\colhead{L$_{FIR}$\tablenotemark{b} } & \colhead{No. of ULX}
}

\startdata
NGC247      & 7.93	& 27.32		& 0.602	& 0.687 & 1	\\
NGC253      & 998.73	& 1861.67	& 55.92	& 93.10	& 3	\\
NGC300      & 23.08	& 74.45		& 1.688	& 1.324	& 0	\\
NGC625      & 5.09	& 9.08		& 0.280	& 0.230	& 0	\\
NGC1313     & 35.97	& 92.00		& 2.329	& 4.845	& 2	\\
IC0342      & 255.96	& 661.68	& 16.66	& 30.32	& 3	\\
NGC1569     & 45.41	& 47.29		& 2.072	& 0.635	& 0	\\
NGC1705     & 0.970	& 2.580		& 0.064	& 0.199	& 0	\\
MRK 71      & 3.51	& 4.67		& 0.173	& 0.239	& 1	\\
NGC2403     & 51.55	& 148.49	& 3.547	& 5.378	& 2	\\
Holmberg II & 1.15	& 2.62		& 0.070	& 0.061	& 1	\\
Holmberg I  & \nodata	& \nodata	& \nodata	& \nodata	& 1	\\ 
M81         & 44.73	& 174.02	& 3.647	& 5.655	& 1	\\
M82         & 1271.32	& 1351.09	& 58.35	& 106.2	& 1	\\
Holmberg IX & \nodata	& \nodata	& \nodata	& \nodata	& 1	\\
Sextans A   & 0.255	& 0.674		& 0.017	& 0.004	& 0	\\
IC 2574     & 2.41	& 10.62		& 0.212	& 0.329	& 0	\\
NGC 4214    & 17.87	& 29.04		& 0.947	& 0.826	& 1	\\
NGC 4258    & 21.60	& 78.39		& 1.690	& 10.48	& 1	\\
NGC4395     & 4.21	& 12.90		& 0.299	& 0.573	& 1	\\
NGC4449     & 37.00	& 58.28		& 1.937	& 2.199	& 1	\\
NGC4490     & 47.79	& 85.94		& 2.636	& 19.19	& 5	\\
NGC4631     & 82.90	& 208.66	& 5.324	& 35.83	& 1	\\
NGC4736     & 62.41	& 135.34	& 3.734	& 8.261	& 4	\\
NGC4945     & 588.11	& 1415.5	& 36.95	& 42.49	& 0	\\
NGC 5204    & 2.33	& 5.35		& 0.143	& 0.395	& 2	\\
M51         & 108.68	& 292.08	& 7.213	& 44.74	& 5	\\
M83         & 266.03	& 638.63	& 16.69	& 76.79	& 0	\\
NGC5253     & 30.00	& 30.92		& 1.365	& 16.72	& 0	\\
M101        & 88.04	& 252.84	& 6.048	& 39.63	& 4	\\
NGC5408     & 2.825	& 2.958		& 0.129	& 0.356	& 1	\\
Circinus    & 248.7	& 315.85	& 12.06	& 23.10	& 4	\\
\enddata

\tablenotetext{a}{flux in units of $10^{-9}$\,erg\,cm$^{-2}$\,s$^{-1}$}
\tablenotetext{b}{far-infrared luminosity in units of 10$^{42}$\,erg\,s$^{-1}$}
\end{deluxetable}

\end{document}